\documentclass[prd,preprintnumbers,floatfix,superscriptaddress,nofootinbib,10pt]{revtex4}
\usepackage{graphicx}
\usepackage{epsfig}
\usepackage{amsmath}
\usepackage{color}
\usepackage{ulem}

\newcommand{\be}{\begin{equation}}

\newcommand{\ee}{\end{equation}}
\newcommand{\ba}{\begin{eqnarray}}
\newcommand{\ea}{\end{eqnarray}}

\begin{document}
\title{Phonon contribution to the shear viscosity
  of a superfluid  Fermi gas in the unitarity limit}
\author{Massimo Mannarelli}
\affiliation{ I.N.F.N., Laboratori Nazionali del Gran Sasso, Assergi (AQ), Italy}
\author{Cristina Manuel}
\affiliation{Instituto de Ciencias del Espacio (IEEC/CSIC) \\
Campus Universitat Aut\` onoma de Barcelona,
Facultat de Ci\` encies, Torre C5 \\
E-08193 Bellaterra (Barcelona), Spain}
\author{Laura Tolos \footnote{e-mail:tolos@ice.csic.es\\ Fax: +34 93 581 43 63, Phone:  +34 93 581 43 65} }
\affiliation{Instituto de Ciencias del Espacio (IEEC/CSIC) \\
Campus Universitat Aut\` onoma de Barcelona,
Facultat de Ci\` encies, Torre C5 \\
E-08193 Bellaterra (Barcelona), Spain}
\affiliation{Frankfurt Institute for Advanced Studies, \\
Johann Wolfgang Goethe University, Ruth-Moufang-Str.~1,
60438 Frankfurt am Main}
\preprint{}

\begin{abstract}
We present a detailed analysis of the contribution of small-angle Nambu-Goldstone boson (phonon) collisions to the shear viscosity, $\eta$, in a superfluid atomic Fermi gas close to the unitarity limit. We show that the experimental values of the shear viscosity coefficient to entropy ratio, $\eta/s$, obtained at the lowest reached temperature  can be reproduced assuming that  phonons give the leading contribution to $\eta$. The phonon contribution is evaluated considering $1 \leftrightarrow 2$ processes and taking into account  the finite size of the experimental system.  In particular, for very low temperatures, $T \lesssim 0.1 T_F$, we find that phonons are ballistic and  the contribution of phonons to the shear viscosity is determined by the processes that take place at the interface between the superfluid and the normal phase. This result is independent of the detailed form of the phonon dispersion law and  leads to two testable predictions: the  shear viscosity should  correlate with the size of the optical trap and it should decrease with decreasing temperature.  For higher temperatures  the detailed form of the phonon dispersion law becomes relevant and, within our model, we find that the experimental data for $\eta/s$   can be reproduced assuming that phonons have an anomalous dispersion law.

\end{abstract}

\maketitle

\section{Introduction}

The study of the  transport coefficients of a fluid  opens a window on its microscopic dynamics, helping us to understand which are the underlying  degrees of freedom and the most important interaction channels \cite{Schafer:2009dj,Adams,Giorgini:2008zz}. Of extraordinary interest is the study of the transport properties of fluids with an infinite two-body scattering length  (corresponding to the so-called unitarity limit), because the expression of the transport coefficients in terms of  the thermodynamic variables is believed to be universal~\cite{Ho:2004zza}, meaning that it is independent of the detailed form of the inter-particle potential. 

Experiments with trapped cold atomic gases are able to reach the region of infinite scattering length. When these experiments are performed with fermionic  atoms, like $^6{\rm Li}$ or $^{40} {\rm K}$, 
prepared in two different  hyperfine states  having a magnetic-field Feshbach resonance,  the strength of the interaction between atoms with opposite spins can be varied  by tuning the  intensity of the magnetic field. 
At very low temperature, by varying the magnetic-field controlled interaction,  fermionic pairing 
is observed to undergo the Bose-Einstein condensate (BEC)  to Bardeen-Cooper-Schrieffer (BCS)  transition, allowing us to explore  the crossover region in a controlled way \cite{Ohara:2002,Bourdel:2003zz,Gupta:2003,Regal:2003}.  Given the adjustability  of the inter-particle interaction,   ultracold fermionic systems can serve to emulate non-relativistic systems (which have the same spontaneously broken symmetries of the ultracold fermionic  system) and  to  determine the universal expression of the transport coefficients at unitarity. On a more general ground, the results obtained  can be used to emulate the properties of different systems which are believed to have similar properties as those of the Fermi gas at unitarity. 
One remarkable example is the fluid realized in relativistic heavy-ion collisions, where  almost ideal hydrodynamic flow has been measured \cite{Adams}.

Cold atomic systems at unitarity can also be used to explore the conjecture  derived by string theory methods \cite{hep-th/0104066}, of the existence of a universal bound of the shear viscosity to entropy ratio $\eta/s \geq  \hbar/ ( 4 \pi k_B)$. 
The fluid produced in heavy-ion collisions seems to have extremely small shear viscosity~\cite{Schafer:2009dj}, close to the universal bound, while it is not yet clear which is the lower bound of $\eta/s$ in  fermionic cold atoms. As we shall see in detail,  the present experimental results give values of $\eta/s$ above the universal bound, but the data seem to indicate  that this quantity would further  decrease  with decreasing temperature. Recently, other different lower bounds on $\eta/s$ have been predicted based on the study of the effect of hydrodynamical fluctuations  \cite{Schafer12,Romat12}.

In the present paper we focus our attention to the superfluid phase of the ultracold Fermi liquid at unitarity. In superfluid  Fermi systems the low energy degrees of freedom are the  Nambu-Goldstone bosons (NGBs),  which arise from the quantum condensation of difermions by a process that spontaneously breaks a global continuous symmetry of the system.  The property of superfluidity follows from the fact that the NGBs at low energy have  a linear dispersion law. We will refer generically to these modes as superfluid phonons, or phonons for simplicity.  

Present experimental data \cite{Ku} show that in the superfluid phase the entropy per particle is extremely small: at the lower reached temperature, $T\simeq 0.1 T_F$, it  is about $0.04 k_B$,   but with rather large error bars, of order $0.1 k_B$\footnote{private communication with Martin Zwierlein}.  These results do not exclude the  possibility that already at this temperature phonons are the dominant degrees of freedom. Although there is no firm experimental result, we shall assume that at such temperature phonons give a sizable contribution to the thermodynamic and transport properties of the system. Further decreasing the temperature there is little doubt that phonons would become  the only relevant degrees of freedom, as the fermions are gapped and  thermally suppressed.
Actually, the Quantum Montecarlo simulations of Ref.~\cite{Bulgac:2008zz}, show that  already  at $T=0.25 T_F$, the total energy, chemical potential and  entropy deviate from the Fermi gas behavior (see Fig.7 of Ref.~\cite{Bulgac:2008zz}). Moreover,   the fact that the chemical potential at very low temperature increases with increasing $T$ is an hint in favor of the contribution of phonons.

In the temperature regime in which  phonons are the only relevant degrees of freedom,  the transport properties of the system  depend on the phonon self-interactions and are sensitive to the phonon Lagrangian. In this case  it is possible to have hints on the phonon dispersion law and on the most important interaction channel by appropriate experimental measurements. In one of his pioneering works,  Landau  proposed a particular phonon spectrum \cite{IntroSupe}, in order to explain the experimental values of the thermodynamic functions of $^4$He. An analogous low energy spectrum can  be inferred from the study of the transverse phonon relaxation processes, see e.g. \cite{Kosevich}. Also 
the most  relevant phonon self-interactions valid for $^4$He can  be determined in a similar way \cite{Maris,Benin}. 

More recently it has been realized that the effective field theory (EFT) techniques can be  used to determine the phonon dispersion law as well as the main phonon self-interactions for many superfluids~\cite{Son:2005rv, valle}.  The EFTs are very efficient  when there is a hierarchy of widely separated energy or momentum scales (in our system, the phonon momentum versus the Fermi momentum), which allows one to obtain an expansion of the effective  phonon Lagrangian  in powers of energy or momentum, rather than in a coupling constant. Power counting and symmetry considerations fix the form of the EFT to the accuracy one desires, regardless of whether the underlying system is weakly or strongly coupled. At leading order in a momentum expansion, one can see that the EFT phonon Lagrangian is related to Landau's Hamiltonian by a Legendre transformation 
\cite{Son:2005rv}. Using the power of these EFT one can thus relate the physics of  phonons in different superfluid systems having the same global symmetries.

In this manuscript  we give an extended explanation of the results of Ref.~\cite{Mannarelli:2012su}, where
we employed both the EFT techniques as well as the experimental available data, to pin  the phonon Lagrangian down. In our procedure we assume a generic expression of the phonon effective Lagrangian, obtained by  an expansion in powers of the phonon momentum, and  study the contributions of various  phonon processes to the shear viscosity coefficient of a superfluid ultracold Fermi liquid  at unitarity.  For the evaluation of $\eta$, we consider the Beliaev  $1 \leftrightarrow 2$ processes, using for this purpose the scattering rates obtained from the EFT Lagrangian of Ref.~\cite{Son:2005rv}. We perform this calculation assuming an infinite volume and employing the results for  the phonon spectrum obtained by various techniques \cite{Rupak:2008xq, Salasnich}. Our results, and associated discussion, can be easily related to those obtained for $^4$He in the temperature, $T$, regime where the viscosity is dominated by the phonons \cite{Maris,Benin}. The resulting expression of the shear viscosity coefficient depends on one free parameter, which is related to the phonon dispersion law at high momentum and that has not been yet evaluated. Then, we try to fix this free parameter by comparing our results with those obtained by the Duke group ~\cite{kinast1, kinast3}, where measurements of the entropy of ultracold fermionic systems and of the shear viscosity coefficient (by measuring the damping of the breathing mode) have been obtained. Unfortunately reaching temperatures much smaller than the Fermi temperature, $T_F$, is extremely difficult and only few experimental points are available at the temperature where the contribution of phonons might be relevant. Moreover, the experimental setup is such that finite size effects must be included for an appropriate comparison with our results.  Indeed, for  $T\simeq 0.1 \,T_F$, we find that for the particular experimental setup of Refs.~\cite{kinast1, kinast3} phonons are ballistic, because  their mean free path is of the same order of the size of the trapped cloud. Thus,  the infinite volume limit we used in the evaluation of the shear viscosity coefficient and the whole hydrodynamical treatment of the phonons ceases to be valid.   However, in the ballistic regime the collisions of  phonons with the boundary still produce dissipation, and ultimately, the damping of the breathing mode. This dissipation can  be described by an effective ballistic shear viscosity, employing an approach  very similar  to the one used to describe the
dissipation in the ballistic regime of the phonons of $^4$He, or other systems  \cite{Cercignani, Sone, Struchtrup}.

Once the parameters of the effective Lagrangian of phonons are known, they can be used  to predict the behavior of different transport properties  at very low temperature, such as the thermal conductivity \cite{Braby:2010ec}, or the bulk viscosity coefficients \cite{Escobedo:2009bh}. In the present paper we content ourselves to consider the behavior of $\eta/s$ at very low temperature, marginally reached by today experiments, and to predict the behavior of this quantity with decreasing temperature.  Indeed,  according to our calculations, trapped phonons below  $\sim 0.1 \, T_F$ are ballistic, and it follows that $\eta/s$ should be directly proportional to  the temperature. We also find that  $\eta$  should approximately scale with the size of the trap, a behavior that   future experiments should be  able to check as well.

This paper is structured as follows. In Sec.~\ref{Inter} we present a brief review of the phonon EFT at leading and next-to-leading order in the momentum expansion, with emphasis on the expression of the phonon dispersion law. We use the EFT to get the scattering rates necessary for 
the evaluation of the contribution to $\eta$ of small-angle collisions in Sec.~\ref{scattrates}. In Sec.~\ref{hydro} we use a variational approach to solve the  Boltzmann equation for three phonon collisions, and obtain their contribution to  $\eta$. The $T$ dependence of such a contribution strongly depends on the form of the phonon dispersion law, and we obtain different values of $\eta$ for some specific choices of the dispersion law in Sec.~\ref{subsec-values}. In Sec.~\ref{experim-eta}
 we discuss the experimental settings of the trapped Fermi gases where the measurements of $\eta$ have been done. In Sec.~\ref{finitesection} we compute the phonon mean free path and compare it with the size of the atomic cloud, to reach to the conclusion that finite size effects have to be taken into account. We write a phenomenological formula for $\eta$ that takes into account those effects, and use it in Sec.~\ref {sec-exp} to give account of the experimental measures of $\eta/s$.
Appendix \ref{appendix-disp-law} discusses why it is enough to keep the lower order corrections  in the phonon dispersion law in the evaluation of $\eta$, while in Appendix \ref{ort-poly} we present an alternative  variational solution to the Boltzmann equation. 
Throughout the paper we use natural units  $\hbar = k_B=1$.

%%%%%%%%%%%%%%%%%%%%%%%%%%%%%%%
\section{Phonon Collisions relevant for the shear viscosity}
\label{Inter}
%%%%%%%%%%%%%%%%%%%%%%%%%%%%%%

We call the superfluid phonon  the  NGB associated to the spontaneous symmetry breaking
of a global $U(1)$  symmetry  group  associated to  particle number conservation.
The phonon field can be viewed, by a  microscopic approach, as the phase oscillation of the difermion condensate when the fermionic degrees of freedom have been integrated out. Actually, in this perspective, the low energy spectrum contains also the  degree of freedom associated to the  oscillation of the absolute value of the difermion condensate, see e.g. \cite{Engelbrecht:1997zz, Gubankova:2008ya, Diener:2008}. But considering only gaussian fluctuation, the radial degree of freedom can be integrated out  and only results in a modification of the phonon dispersion law. In the following we assume that in the EFT Lagrangian the radial oscillations of the condensate have already been integrated out and  the phonon is  the only dynamical degree of freedom. It is worth to note that the fluctuations in the absolute value of the condensate are not taken into account in the analysis of \cite{Salasnich-2,Combescot:2006zz} where the Lagrangian of the collective modes has been determined starting from a mean-field Hamiltonian.

 Effective field theory techniques can be used to derive the Lagrangian of the superfluid system at unitarity by a derivative expansion in the  phonon field. At the lowest order,  the effective Lagrangian has been obtained in  Ref.~\cite{Son:2005rv}
and is formally given at vanishing temperature by the functional
\begin{equation}
\label{LO-Lagran}
\mathcal{L}_{\rm LO} =P (X) \ ,
\end{equation}
with 
\begin{equation}
\label{Xvariable}
 X = \mu_0- V({\bf r}) - \partial_t\varphi-\frac{({\bf \nabla}\varphi)^2}{2m} \, ,
\end{equation}
where $P(\mu_0)$ and $\mu_0$ are the pressure and chemical potential, respectively, of the non relativistic superfluid at $T=0$, 
$\varphi$ is the phonon field, and $V$ is the trapping potential.  For $V=0$, the equation of state (EoS) of the unitary Fermi gas reads
\be
P = \frac{2^{5/2}}{15 \pi^2 \xi^{3/2}} m^{3/2} \mu_0^{5/2} \,,
\label{pressure}
\ee
where  $m$ is is the mass of the particles that condense and  $\xi \simeq 0.36-0.38$ \cite{Ku,  Haussmann:2007zz, Arnold:2006fr}  is the Bertsch number \cite{Bertsch}, a universal constant  that fixes the relation between  chemical potential and  Fermi energy, $\mu_0 = \xi E_F$.

 It is convenient to re-express the LO Lagrangian in a different way. After a derivative expansion, and rescaling of the phonon field to have a canonically normalized kinetic term, one obtains 
\begin{equation}
\label{comlag}
\mathcal{L}_{\rm LO}=\frac{1}{2}\left((\partial_t\phi)^2-v^2_{\rm ph}({\bf \nabla}\phi)^2\right)-g\left((\partial_t \phi)^3-3 g_s \,\partial_t \phi({\bf \nabla}\phi)^2 \right)
+ {\cal O} ( (\partial \phi)^4)
+ \cdots \ .
\end{equation}
where we have set $V=0$ for simplicity. The different  self-coupling constants of Eq.~(\ref{comlag}) can be  expressed
in terms of derivatives of the pressure with respect to the chemical potential \cite{Escobedo:2010uv}. For comparison
with the condensed matter literature on superfluidity, and for the purposes of computation,
it turns out to be more convenient to express them in terms of the density $\rho$, the speed of sound at $T=0$, $c_s$, and derivatives
of the speed of sound with respect to the particle density (see Appendix A of Ref.~\cite{Escobedo:2010uv})
\begin{equation}
%\label{phspeed}
v_{\rm ph}=  c_s= \sqrt{ \frac 1m \frac{\partial P}{\partial { \rho}} } \ , \qquad
g=\frac{1}{6 \sqrt{m\rho} \ c_s } \left(1-2 u \right) \ , \qquad
g_s = \frac{c_s^2}{1-2 u \ }  \ ,
 \label{eq:g}
 \ee
where $u =\frac{\rho}{c_s}\frac{\partial c_s}{\partial \rho}$ is the so-called Gr\"uneisen number.

The next-to-leading order (NLO) Lagrangian is constructed
 by demanding non-relativistic general coordinate invariance and conformal symmetry~\cite{Son:2005rv}, and it reads
\be
\label{effL-LO+NLO}
{\cal L}_{\rm NLO} = c_1 m^{1/2} \frac{(\nabla X)^2}{\sqrt{X}} + \frac{c_2}{\sqrt{m}}(\nabla^2 \varphi)^2  \sqrt{X} \,,
\ee
 where $c_1$ and $c_2$ are two  dimensionless parameters, which are universal  but cannot be determined by symmetry considerations.   Both coefficients determine not only the corrections to the different phonon self-couplings in Eqs.\eqref{eq:g}, but, and this  is more important for the evaluation of the shear viscosity,  also  the corrections to the phonon dispersion law which at the NLO   reads~\cite{Son:2005rv}
\be
\label{NLOdisp-law}
E_k = c_s k (1 + \gamma  k^2 )  \ ,
\qquad
\gamma = -\left(c_1 + \frac{3}{2}c_2\right) \frac{\pi^2 \sqrt{2 \xi}}{k_F^2}  \ ,
\ee
where $k_F = \sqrt{2m E_F}$ is the Fermi momentum.  The sign of $\gamma$ has a dramatic effect on the possible phonon interaction channels, because, as we shall discuss in detail in the next section, the Beliaev process $1 \leftrightarrow 2$ between massless particles is only allowed for  positive values of $\gamma$. The numerical values of $c_1$ and $c_2$ are still uncertain, but some progress has been made by a number of different techniques. Both coefficients can be estimated by the $\epsilon$-expansion technique, which gives $c_1+3/2 c_2 \simeq -0.0209$ \cite{Rupak:2008xq}. However,  these estimates include large errors, in particular for the determination of $c_2$, because a finite value of  $c_2$ only appears at the second order in the $\epsilon$ expansion \cite{Rupak:2008xq}. A different approach is presented in   \cite{Salasnich} where the phonon dispersion law is determined   by a fit of Quantum Monte-Carlo (QMC) simulations. The reported result is $c_1+3/2 c_2 \simeq -0.028$, but large uncertainties are present, arising from the numerical fit of the  QMC data.  A mean-field theory has been employed in Ref.~\cite{valle}, giving  $c_1+3/2 c_2 \simeq -0.0205$.  Although with large numerical uncertainties all these methods give results  consistent with a positive  $\gamma$ coefficient.

It is of some relevance the fact that the NLO Lagrangian in Eq.~\eqref{effL-LO+NLO} might not be the correct starting point for the description of the superfluid phonon field. According to  \cite{valle}, the reason is that such NLO Lagrangian has been obtained by a field redefinition which changes the NLO meaning of the phonon field. In other words, the field $\phi$ appearing in Eq.~\eqref{effL-LO+NLO} through the field $X$, is not related to the superfluid velocity by the standard relation, thus it is not true that at the NLO order $v_s \propto \nabla \varphi$, and therefore $\phi$ cannot be interpreted as the superfluid phonon. If the expression of the NLO Lagrangian of \cite{valle},  containing an additional term, is used, it leads to a change of the expression of the NLO contribution to the dispersion law as in Eq.~\eqref{NLOdisp-law}, but with $\gamma \propto - (c_1-3c_2)$, and thus still making $\gamma$  positive.
  
A different estimate of these parameters can be obtained by means of a mixed approach \cite{Haussmann:2009}, which combines the $\epsilon$-expansion technique and the results obtained considering  the gaussian fluctuations on the top of the mean field solution \cite{Diener:2008}. According to  \cite{Haussmann:2009}, $c_2$ is proportional  to the square of the pair size and should therefore be positive (in disagreement with the estimate of  \cite{valle} ) and presumably comparable in size to $c_1$. The results of \cite{Haussmann:2009}   lead to $c_1+3/2 c_2 \simeq  + 0.01$ and therefore to a negative value of $\gamma$, in disagreement with  \cite{Rupak:2008xq, Salasnich, valle}. However, considering gaussian fluctuations close to the unitarity might not be enough to determine the dispersion law of the collective modes, because the procedure seems to be not self-consistent \cite{Diener:2008}.  For this reason  we assume that   
\be \gamma  \simeq \frac{ 0.18}{k_F^2}\,,\ee 
which is obtained considering $c_1+3/2 c_2 =  -0.021$ and $\xi=0.37$.  We shall then  briefly comment on how our results would change for negative values of $\gamma$.

Given the expression of $X$ in Eq.\eqref{Xvariable}, it is clear that in the NLO Lagrangian, not only the trapping potential appears, but also its space derivatives.  In order to simplify the calculations we will use  the local density approximation (LDA), which corresponds to consider the system as locally homogeneous,  and therefore we will neglect  the spatial derivatives of the trapping potential. Approximating the  trapping potential  by the harmonic potential
\be 
V({\bf x}) = \frac 12 m\,\omega_i^2 x_i^2\,,
\label{eq:pot}
\ee 
the LDA turns out to be a good approximation as far as the local value of the chemical potential is much larger than  $\omega_i$, which is a condition that is satisfied in the experimental settings of Ref.~\cite{kinast1, kinast3}, except near the edge of the cloud.  In the following we will first consider the situation where there is no trapping potential, $V=0$, and then comment on the effects in the phonon physics due to the presence of the trap.

Terms in the phonon Lagrangian beyond the NLO  will as well change the  phonon dispersion law and the interaction vertices. 
 We are not aware, to the best of our knowledge, of a determination of the phonon dispersion law at higher orders\footnote{See, however, the results of
Ref.~\cite{Salasnich-2}, which refer to the dispersion law of hydrodynamical sound waves.}.  In general,  the  phonon dispersion law can be written as
\be
\label{disp-law}
E_k = c_s k (1 + \psi (k) )  \ ,
\ee 
where  $\vert \psi(k) \vert \ll 1$ and  it  can be  Taylor expanded in $k^2/k_F^2$,
 \be\label{psik}
\psi(k) = \gamma\, k^2 + \delta\, k^4 + {\cal O}\left(\frac{k^6}{k_F^6}\right)\,,
\ee 
where $\delta \propto 1/k_F^4$ and  we shall neglect terms of order higher or equal to $ k^6/k_F^6$. We shall obtain a numerical estimate of $\delta$ from a fit of the experimental values of the shear viscosity to entropy ratio. In doing this we shall follow a procedure akin to the one used in $^4$He, where the spectrum of the low energy excitations is determined from a fit of the experimental values of the mean free path of phonons, see e.g. \cite{Kosevich}.  We comment on the appropriate temperature range of  validity of the expansion in Eq.\eqref{psik} in the Appendix \ref{appendix-disp-law}, where also an estimate of the $k^6/k_F^6$ term is given.

In principle, from a naive power counting, one would  expect that NLO (or higher order) corrections to the interaction vertices  should be taken into account as well. However, we shall show that in the computation of the shear viscosity coefficient these corrections are subleading with respect to the one determined by the dispersion law. The reason being that, as we shall discuss in detail in the next section, the dispersion law determines the  power dependence of the shear viscosity coefficient on the temperature. On the other hand,  the neglected terms in the interaction vertices do always give  corrections of higher power in $T/T_F$.

\subsection{Three phonon and four phonon interactions}
\label{scattrates}

For the computation of the shear viscosity coefficient one needs to evaluate  the scattering rates of both the binary (hereafter 4-ph) collisions and the Beliaev $1  \leftrightarrow 2$ (hereafter 3-ph) processes. 
The vertices associated to the corresponding interactions can be obtained by the momentum expansion of the Lagrangian  in Eq.\eqref{LO-Lagran},  and then the scattering amplitudes are readily obtained by standard techniques. 

Both the contributions of 3-ph and 4-ph processes to the shear viscosity  depend on the dispersion law of the phonons. However, the 3-ph processes are more sensitive to the expression of $\psi(k)$ because of kinematical restrictions.
Considering  the LO  phonon dispersion law, $E_p = c_s p$, one finds that,  for 3-ph processes, energy and momentum conservations  only allow the collinear scattering, and  the contribution of this process to the shear viscosity diverges. The reason being that the shear viscosity coefficient is proportional to the time scale for momentum transfer in the orthogonal direction to the flow, and if the collision is collinear it takes an infinite amount of time to transfer momentum in the orthogonal direction.

Considering the generic corrections to the LO Lagrangian, the dispersion law has an additional term $\psi(k)$,  see Eq.~\eqref{disp-law}, which may favor or disfavor $1 \leftrightarrow 2$ processes depending on the sign of $\psi(k)$. If $\psi(k) <0 $, 3-ph processes are kinematically forbidden.  On the other hand, if    $\psi(k) >0$ (at least for some momenta $k$),  3-ph processes are kinematically allowed. In the following we shall assume that the NLO corrections give $\gamma >0$ and therefore make $\psi(k)$ positive for certain values of $k$.

If the three phonons participating in the collision have momenta  ${\bf p}$, ${\bf p'}$ and ${\bf k}$, then the collision angle between the phonons with momenta ${\bf p}$ and ${\bf p'}$ is given by
\be
\label{coll-angle}
\cos{\theta_{p p'}} = 1 +  \frac{1}{p'} \left(p' \psi(p') + k' \psi(k') -(p'+k') \psi(p'+k') \right) \,, 
\ee  
and since $\psi(k) \ll 1$, one can deduce that the deflection angle is small, and roughly proportional to  $\theta \sim \psi^{1/2}$. After an analysis of the scattering processes one obtains that phonons with thermal momentum
close to  $k_{\rm th} \sim 10 T/c_s$ give the leading contribution to the viscosity, and  one can estimate   $\theta_{\rm th} \sim \psi_{\rm th}^{1/2}$, where $\psi_{\rm th}=\psi(k_{\rm th})$. 
 
In  the following analysis of the contribution of the 3-ph processes to the shear viscosity,  we will be using a double expansion, both in the phonon momentum and in the collision angle.
 We will be considering the vertices of the scattering rates as arising from the lowest momentum expansion ${\cal L}_{\rm LO}$, as corrections from ${\cal L}_{\rm NLO}$
would be suppressed  in the shear viscosity by powers of  $T/(c_s k_F) \propto T/T_F $. However, we will use the phonon dispersion law beyond LO, as the expression of $\psi(k)$ determines the collision angle,
and approximate all different expressions in small angles.

From the LO Lagrangian we have that the square of the  scattering amplitude of the 3-ph processes   is given by
\be
\label{scat-mat}
|{\cal M}_{\rm 3ph}|^2= 36 g^2 \left( - E_p E_{p'} E_{k'} + g_s \left( E_p {\bf p' } \cdot {\bf k' } + E_{p'} {\bf p } \cdot {\bf k' } +E_{k'} {\bf p } \cdot {\bf p' } 
\right) \right)^2 \ , 
\ee
where $g_s$ is defined in Eq.~(\ref{eq:g}), $E_p, E_{p'}$ and $E_{k'}$ are the energies of the phonons with momenta  ${\bf p}, {\bf p'}$ and ${\bf k'}$, respectively. After using the constraints of energy and momentum conservation and considering phonons with the dispersion law in Eq.~(\ref{disp-law}), the LO scattering amplitude can be expressed as
\be\label{approxM}
|{\cal M}_{\rm 3ph}|^2= \frac{4 c_s^4  \left( 1+ u \right)^2}{\rho}( p p' k' )^2 + {\cal O}(\psi) \ .
\ee
This expression is  formally equivalent to  the scattering amplitude for the 3-ph processes in superfluid
$^4$He, and it has been used by Maris \cite{Maris} and Benin \cite{Benin} for the computation of the shear viscosity in that system.
What changes are  the explicit expressions of $\rho$ and $c_s$  as well as the ${\cal O}(\psi)$ corrections.
The fact that the expressions of the 3-ph processes are the same in these two superfluid systems
is not unexpected, as it was shown \cite{Son:2005rv} that the formal expression of the effective Lagrangian at LO, Eq.~(\ref{LO-Lagran}),
can be related by a Legendre transformation to the Hamiltonian proposed by Landau to derive the phonon self-interactions in superfluid $^4$He.

The expression of the scattering amplitude of the 4-ph process has been derived at the LO in momentum expansion  in Ref.~\cite{Manuel:2011ed}. We refer the reader to that paper for the corresponding expression of $|{\cal M}_{\rm 4ph}|^2$. 

The results obtained above still hold when including the trapping potential in the local density approximation.  The phonon effective field theory, and the associated scattering amplitudes of the 3-ph and 4-ph processes,  are  modified by taking into account that the $x$-dependence of the trapping potential can be included in a  space dependent effective chemical potential, $\mu ({\bf x}) = \mu_0 -V({\bf x})$.  But this is not the only effect due to the presence of the trap. As we shall discuss in detail in Sec.~\ref{experim-eta}, the restricted geometry dictated by the trapping potential leads to important boundary effects when the mean free path of phonons becomes comparable with the typical length scale of the trap.

\section{Shear viscosity due to small-angle collisions}
\label{Shear-sec}

The shear viscosity $\eta$  measures the transport of momentum in the orthogonal direction to the hydrodynamic flow.
It is generally dominated by large-angle collisions of the quasiparticle components of the fluid. There are
cases, however, when small-angle collisions may dominate, 
as it  may turn out to be more efficient to achieve a large-angle collision by the addition of many small-angle scatterings. In the low $T$ regime of superfluid $^4$He where the viscosity is dominated  by phonons, the 4-ph 
large-angle collisions give the leading contribution to the shear viscosity only in a restricted range of temperature \cite{Zadorozhko}, while 3-ph small-angle processes dominate for any temperature below $0.7\,$K \cite{Maris}.

In this section we use a kinetic theory approach to compute the 3-ph contribution to the shear viscosity, assuming that
the system is infinite and homogeneous. We use the scattering rate as arising from ${\cal L}_{\rm LO}$ evaluated in the previous section, but assume in the energy and momentum conservation the generic phonon dispersion in Eq.\eqref{disp-law}, with $\psi(k)>0$. We shall discuss in detail the range of validity of this method, that is why in the scattering rate the ${\cal O}(\psi(k))$ corrections can be neglected. Eventually, we perform both an expansion in the phonon momentum and in the collision angle given in  Eq.~(\ref{coll-angle}),  finding that the temperature dependence of the shear viscosity  strongly depends on the expression of the phonon dispersion law.

%%%%%%%%%%%%%%%%%%%%%%%%%%%%%%%%%%%%%
\subsection{Variational solutions of the Boltzmann equation}
\label{hydro}

%%%%%%%%%%%%%%%%%%%%%%%%%%%%%

When a  shear stress is  applied to a fluid, e.g. by contact with a sliding surface, it will perturb the fluid producing a gradient of the velocity field. In a superfluid, a sliding surface moving at sufficiently small velocity will only produce a gradient of the normal field velocity  in the direction perpendicular to the  surface. For small deviations from equilibrium, the energy-momentum tensor is given by
\be
\label{shear-stress}
 \delta T_{ij}=- \eta  V_{ij}  \equiv  - \eta\left( \partial_i V_j+ \partial_j V_i -\frac 23 \delta_{ij} \nabla \cdot  {\bf V} \right) \ , 
\ee
where ${\bf V}$ is the velocity of the normal fluid component. 

At the microscopic level, the  phonon contribution to  the energy-momentum tensor of the system can be computed by means of  kinetic theory and it is expressed by
 \be\label{kin-Tij}
 T_{ij}= c_s^2 \int \frac{d^3 p}{(2 \pi)^3}  \frac{ p_i p_j}{E_p} f(p,x)  \ , 
  \ee
where $f(p,x)$ is the phonon distribution function, which obeys the Boltzmann equation \cite{IntroSupe}
 \be
  \label{transport}
   \frac{df}{dt} = \frac{\partial
f}{\partial t}+ \frac{\partial E_p}{\partial \bf p} \cdot \nabla f= C[f] \ ,
\ee 
where we have assumed to be in the superfluid rest frame, and $C[f]$ is the collision term. The processes that give the largest contribution to the collision term are  4-ph collisions and 3-ph splitting and joining processes. We shall assume that these processes can be evaluated separately (i.e. independently) and then compare the corresponding equilibration times. 
The contribution of the binary collisions to the shear viscosity has already been evaluated in Refs.~\cite{Rupak:2007vp,Manuel:2011ed}. Therefore, here we shall focus on the 3-ph processes. For the $1 \leftrightarrow 2$ processes  the collision integral is given by
\begin{eqnarray}
C_{1 \leftrightarrow 2}[f] & = &  - \frac{1}{4E_p} \int_{p',k'} |{\cal M_{\rm 3ph}}|^2  (2\pi)^4 \delta^{(4)}( P-P'-K')
\left \{ f(p)  (1+f(p'))(1+f(k')) - f(p')f (k')(1+f(p))\right \} \nonumber
 \\
&+&  \frac{1}{2 E_p} \int_{p',k'} |{\cal M_{\rm 3ph}}|^2  (2\pi)^4 \delta^{(4)}( P'-P-K')
\left\{f(p')  (1+f(p))(1+f(k')) - f(p) f(k')(1+f(p'))\right\} \ ,
\end{eqnarray}
where we have defined the shorthand notation 
\be
\int_p \equiv \int  \frac{d^3 p}{(2 \pi)^3 2E_{p}} \ ,
\ee
and the square of scattering amplitude $ {\cal M_{\rm 3ph}}$ was written in Eq.~(\ref{scat-mat}). As expected,
the collision term vanishes when evaluated using the phonon Bose-Einstein equilibrium distribution function, so that $C[f_{\rm eq}]=0$.

  For the computation of the transport
coefficients we consider small departures from equilibrium
so that the distribution function can be written as
$ f= f_{\rm eq}+ \delta f $, 
and then we linearize the transport equation in $\delta f$. For the computations of the shear
viscosity, one assumes that 
\be\label{deltaf}
\delta f = - h(p) p_{kl} V_{kl} \frac{f_{\rm eq} ( 1+ f_{\rm eq})}{T}    \ ,
\ee
where $h(p)$ is an unknown function, $V_{kl}$ is the tensor defined in Eq.~(\ref{shear-stress}), and we have defined 
\be
p_{kl} = p_k p_l - \frac 13 \delta_{kl} p^2 \ .
\ee

Upon substituting the expression of the perturbed distribution function in Eq. \eqref{kin-Tij}
one can extract the value
of  the shear viscosity coefficient, which turns out to be
 \be
 \label{1stexpsh}
 \eta_{\rm 3ph}=\frac{4 c_s^2 }{15 T }\int
\frac{d^3p}{(2\pi)^3} \frac{p^4}{2 E_p}  f_{\rm eq}(1+f_{\rm eq}) h(p) \ .
 \ee

On the other hand, inserting Eq.\eqref{deltaf} in the collision integral, we obtain at the linear order 
\ba
\delta C & \equiv & \frac {1}{2E_pT} F_{ij}[h(p)] V_{ij} \ ,
   \ea
and the linearized Boltzmann equation reads
\be\label{boltz-linear}
c_s \frac{ f_{\rm eq}(1+f_{\rm eq})  }{2pT} p_{ij} V_{ij} = \frac {1}{2E_pT} F_{ij}[h(p)] V_{ij} \ .
\ee

With this last result one realizes that the shear viscosity can also be written as
\be
\label{2onexpsh}
 \eta_{\rm 3ph}=\frac{2 }{5 T} \int_p  p_{ij} h(p) F_{ij}[h(p)] \ ,
 \ee
or equivalently,
\be
\label{shear-visco}
\eta_{\rm 3ph} =   \frac{1}{5T}  \int_{p,p',k'}  (2\pi)^4 \delta^{(4)}( P- P' -K')  |{\cal M_{\rm 3ph}}|^2 f^{\rm eq}_{p}  ( 1+f^{\rm eq}_{p'})(1+  f^{\rm eq}_{k'})  \left(\Delta_{ij}^h\right)^2 \ ,
 \ee
where
\be
\Delta_{ij}^h \equiv h(p) p_{ij} - h(p') p'_{ij} - h(k') k'_{ij}  \ ,
\ee
and we have used the property that the scattering amplitude is symmetric in the momenta of the three phonons.

In order to have a lower bound on the shear viscosity coefficient it is possible to employ a variational method. In order to do so, one defines the inner product
\be
\langle \phi_1 \vert \phi_2 \rangle = - \int_p \phi_1(p) \phi_2 (p) \frac{d f_{\rm eq}}{d E_p} \ ,
\ee
where $\vert \phi_1 \rangle$ and $\vert \phi_2 \rangle$ are two arbitrary functions. If we define the function $| \chi \rangle = p_{ij}$,  then Eq.\eqref{boltz-linear}  can be schematically written
as $| \chi \rangle = H | \Phi \rangle $, where $ \Phi$ is the solution of the linearized transport equation.
 The shear viscosity can be written as  $\eta_{\rm 3ph} =   \langle \chi | \Phi \rangle $, which is equivalent to  Eq.\eqref{1stexpsh}, or alternatively,   $\eta_{\rm 3ph} =   \langle \Phi | H | \Phi \rangle$, as in Eq.\eqref{shear-visco}. Finally,  from the Schwarz inequality (and considering that $H$ is positive semidefinite)  it follows that for an arbitrary function $\tilde \Phi$,
\be\label{variational}
  \eta_{\rm 3ph} \geq \frac{  \langle \chi | {\tilde \Phi} \rangle^2  }{  \langle {\tilde \Phi} | H | {\tilde  \Phi} \rangle}\, ,
 \ee
which is saturated when ${\tilde \Phi}$ is the exact solution of the linearized transport equation. 

We will compute the value of $\eta_{\rm 3ph}$ using this variational method, and assuming that the phonon dispersion law is of the form given in Eq.~(\ref{disp-law}), where $\psi(k)$ is treated as a small perturbation. 
In principle an upper bound of the shear viscosity can as well be obtained,  e.g. by the method proposed in  \cite{Hojgaard2}. However,   from the fact that $\eta_{\rm 3ph}$ should diverge for $\psi(k)=0$, it is possible  to figure out the  particular  family of functions which maximizes Eq.\eqref{variational}.
More in detail,  we look for  a solution that minimizes $(\Delta_{ij}^h)^2$, and thus, considering the expression of $H$, maximizes the value of $\eta_{\rm 3ph}$. It is then easy to see that within the space of rational functions,  the solution must be of the form
\be
\label{trial-a}
h(p) = \frac 1p \left(1 + a \psi(p) \right) + {\cal O} ( \psi^2) \ ,
\ee
where $a$ is the variational parameter. This form of the solution is such that $(\Delta_{ij}^h)^2 \propto \psi^2$,
while other forms of rational functions give a result of order one, and thus, bigger  in the $\psi$-expansion. 
We fix the value of $a$ by minimizing 
\be
\Delta_{ij}^h \Delta_{ij}^h =  \frac 19 (42+24 a + 6 a^2)  \left( p' \psi(p') + k' \psi(k') - (p'+k') \psi(p'+k') \right)^2\,,
\ee
 which occurs when $a=-2$. We have  checked the stability of this variational solution by changing the form of variational functions used in the maximization procedure, see Appendix~\ref{ort-poly}.
We have also checked that our variational solution agrees with the variational solution of the phonon transport equation given in Ref.~\cite{Benin}, where the shear viscosity due to joining and splitting processes for superfluid  $^4$He has been computed.

With the form of $h(p)$ reported above, we computed the shear viscosity coefficient at the leading order in  $\psi$, and we obtained  
\be
\eta_{\rm 3ph} = \left( \frac{2 \pi}{15}\right)^4 \frac{T^8}{c_s^8}  \frac{1}{M} + {\cal O} ( \psi^2) \ ,
\ee
where
\be
\label{theMinteg}
M =   \frac{   \left( 1+ u \right)^2} {20  T \pi^3 \rho}\int^{*} dp' \,dk'\, (p' k' ( p' +k') )^2  
   \left( p' \psi(p') + k' \psi(k') - (p'+k') \psi(p'+k') \right)^2
f^{\rm eq}_{p'+k'}  ( 1+f^{\rm eq}_{p'})(1+  f^{\rm eq}_{k'}) \ ,
\ee
 and the star on top of the integral indicates that the integration is constrained by the energy and momentum conservation. As already remarked, the expression of the shear viscosity given above strongly depends on the
form of the phonon dispersion law. We will explicitly check this fact in the  following subsection  by computing 
$\eta_{\rm 3ph}$  for different choices of $\psi(p)$.  Notice that in   Eq.\eqref{theMinteg} we did not need to include  in the scattering amplitude  the ${\cal O}(\psi)$ corrections. Actually, in  Eq.~\eqref{approxM} we did not even calculate such corrections. We shall see at the end of the next section that neglecting these correction is equivalent to neglect temperature corrections of order ${\cal O}(T/T_F)^2$ in our final results.

\subsection{Values of the viscosity with different phonon dispersion laws}\label{subsec-values}

Let us first consider a dispersion law such that $\psi (k) = \gamma k^2$ (with $\gamma >0$), meaning that 3-ph processes are allowed for any value of the momentum of the interacting particles. By a power counting analysis one readily  deduces the scaling
$\eta \propto 1/T^5$, which is the same  scaling  obtained with binary collisions of phonons \cite{Rupak:2007vp,Manuel:2011ed}, and from the expressions derived  above we obtain
\be
\label{etagamma}
\eta_{\rm 3ph} \simeq 2.1 \times 10^{-7} \frac{c^6_s  \,\rho}{T^5 \gamma^2  \left( 1+ u \right)^2}  \ .
\ee

Neglecting ${\cal O}(\psi )$ corrections, the entropy due to the phonons is given by
\be\label{entropy}
s_{\rm ph} = \frac{2 \pi^2 T^3}{45 \, c_s^3} \,,
\ee
and assuming the values $\xi = 0.37$, and that $|c_1 + 3 c_2/2| \sim 0.021$, we obtain
\be
\frac{\eta_{\rm 3ph}}{s_{\rm ph}} \simeq 4.0  \times 10^{-9} \frac{T_F^8}{T^8} \,. 
\ee
This should be compared with the shear viscosity coefficient obtained considering  binary collisions of phonons. At the LO in the momentum expansion  \cite{Schafer:2009dj,Rupak:2007vp,Manuel:2011ed}
\be
\label{4pheta}
\frac{\eta_{\rm 4ph}}{s_{\rm ph}} \simeq  2.2 \cdot 10^{-7} \frac{T_F^8}{T^8}  \,,
\ee
which is about two orders of magnitude larger than $\eta_{\rm 3ph}$.  Also notice that $\eta_{\rm 4ph}$ does not diverge for $\gamma =0$ because no kinematical restriction applies to 4-ph processes. 

At this point it is important to remark that the above expressions   lead to two unphysical results. Both $\eta_{\rm 3ph}$ and $\eta_{\rm 4ph}$ diverge at vanishing temperature,  and both are below the experimental values (and  below the universal limit) for sufficiently high temperatures. We shall see that  these behaviors are due to  an oversimplified description of the phonon fluid adopted so far, and they can actually be used to have information about the phonon fluid. We shall discuss the $T \to 0$ limit in Sec. \ref{experim-eta}, and we shall see that the behavior of the shear viscosity in this limit is regulated by the geometrical extension of the trap.

The decrease of $\eta_{\rm 4ph}/s$ with increasing temperature  is due to the fact that a linear dispersion law was used in its computation \cite{Rupak:2007vp,Manuel:2011ed}, while this behavior should be modified when considering higher order terms in the phonon dispersion law. 
Given that  $\gamma \propto 1/k_F^2$ and  treating $\gamma T^2/c_s^2$ as a
perturbative parameter, we expect that the first NLO correction to $\eta_{\rm 4ph}$ leads to
\be
\label{4pheta-n}
\frac{\eta_{\rm 4ph}}{s_{\rm ph}} \simeq  2.2 \cdot 10^{-7} \frac{T_F^8}{T^8} + B  \frac{T_F^6}{T^6}  \,,
\ee
where $B$ is some coefficient that depends linearly on $\gamma$, and thus  vanishes for $\gamma=0$.  The correction becomes more and more important when approaching $T_c$   and moreover with increasing temperature we expect that more terms   should be included in the expansion of $\eta_{\rm 4ph}$ in Eq.\eqref{4pheta-n}, eventually preventing $\eta_{\rm 4ph}/s_{\rm ph}$ to decrease below the universal bound.  Unfortunately, the evaluation of the coefficient $B$ and of the other corrections is rather complicated. For this reason  we shall not include in our discussion the contribution of 4-ph processes.

The decrease of  $\eta_{\rm 3ph}/s$ with increasing temperature is due to the fact we have assumed a dispersion law with $\psi(k) = \gamma k^2$. Since the typical momentum of phonons is of order $10\, T/c_s$, by increasing the temperature, the neglected terms in the $\psi(k) $ expansion may become relevant. This motivates us to include terms of order $k^4/k_F^4$ in the expansion of Eq. \eqref{psik}, that is, we take into account the next to next to leading order  (NNLO) term in the momentum expansion of the dispersion law. While there are different computations in the literature of $\gamma$, the value of $\delta$ has not been computed yet. We only know that 
$\delta \propto 1/k_F^4$. We assume that $\delta <0$, as the opposite choice leads to even higher values
of the viscosity at low $T$ than those obtained in Eq.~(\ref{etagamma}). In the following section  we shall determine the approximate value of $\delta$ from a fit of the experimental data.

With this choice of the sign of $\delta$,   the function $\psi$ is only positive in a restricted region of momenta. This has as a direct consequence that the three phonon processes  are only possible when the phonon momenta are in a certain range of values. The constraints of energy and momentum conservation at the 3-ph vertex, implies that the integration ranges in the integral in Eq.~(\ref{theMinteg}) are given by 
$0\le k^\prime \le \sqrt{\frac{6}{5\beta} }$, where  $\beta =- \frac{2\delta}{\gamma}$, and   by
$p^\prime_1 \le p^\prime \le p^\prime_2$, where
$p^\prime_{1,2}=(-k^\prime \pm \sqrt{-3 k^{\prime 2} + \frac{24}{5 \beta}} )/2$. After defining the dimensionless parameter
\be\label{gamma}
\psi_{\rm max} = -\frac{\gamma^2}{ 4 \delta} \,,  
\ee
and  \be \tilde\beta = -\frac{2 \delta}{\gamma} \frac{T^2}{c_s^2}\,,\ee
we can express the shear viscosity coefficient as
\be\label{eta-3ph}
\eta_{\rm 3ph} =  \frac{2^6  \pi^7 }{3^4 5^3} \frac{c_s^2 \, \rho}{T (1+u)^2} \frac{1}{\psi^2_{\rm max}\, I(\tilde \beta)} \,,
\ee
where 
\be
I(\tilde\beta) = \tilde\beta^2  \int^{*} dx dy (xy (x+y))^4 (6 - 5 \tilde \beta( x^2+y^2+xy))^2 f_{x+y} (1+f_x ) (1+f_y )  \,,
\ee
and  $x= p' c_s/T$ and $y = k' c_s/T$, so that the integral is done over adimensional variables.
  This leads to the following expression of the shear viscosity to entropy ratio (we take  $\xi = 0.37$)
\be
\frac{\eta_{\rm 3ph}}{s_{\rm ph}} \simeq \frac{ 0.09 }{ \psi_{\rm max}^2}  \frac{1}{I(\tilde\beta)} \frac{T_F^4}{T^4} \,.
\label{eta-3phallT}
\ee

The quantity $\tilde\beta$ depends on $T$, and therefore the temperature dependence of $\eta_{\rm 3ph}/s$ is complicated. However, we find that for 
$0.003\lesssim \tilde\beta\lesssim 0.015$,
$I(\tilde\beta) \approx 2046.4$, independent of $T$. In this region, we thus obtain that $\eta_{\rm 3ph} \propto 1/T$, and then 
\be
\frac{\eta_{\rm 3ph}}{s_{\rm ph}} \simeq \frac{4.2~10^{-5}}{ \psi_{\rm max}^2} \frac{T_F^4}{T^4}  \qquad {\rm for}\,\,\, 0.05\,T_F \lesssim T  \lesssim 0.12\,T_F\,,
\label{eta-ph-ph}
\ee
where the temperature range has been evaluated considering  $\psi_{\rm max} \sim 0.2 - 0.4$, see Sec. \ref{sec-exp}.

For smaller values of the temperature we obtain that $I(\tilde\beta)  \propto \tilde\beta^2$, leading to a viscosity $\eta_{\rm 3ph} \propto 1/T^5$, in agreement with the results obtained using a  vanishing $\delta$. This is as expected, because the leading contribution to $\eta_{\rm 3ph}$ comes from phonons with momenta of the order $10 T/c_s$, and  at very low temperature, the correction introduced by the $\delta$ term of the dispersion law is of the order $ T^4/(c_s k_F)^4$, and becomes negligible.
We have not obtained an analytical expression of $I(\tilde\beta)$ for arbitrary values  of ${\tilde \beta}$, thus we will 
present only numerical results in Sec.~\ref{experim-eta} for particular values of  $\psi_{\rm max}$.

From these results it is apparent that the 3-ph contribution to the shear viscosity coefficient is extremely sensitive to the dispersion law of phonons. At the LO (neglecting terms proportional to $\gamma$ and $\delta$), the shear viscosity coefficient diverges (see the discussion after Eq.\eqref{scat-mat}). Including the NLO term proportional to $\gamma$ leads to $\eta_{\rm 3ph} \propto 1/T^5$, and finally including the NNLO term proportional to $\delta$ gives, in a certain temperature range,  $\eta_{\rm 3ph} \propto 1/T$. These results have to  be contrasted with the effect of the neglected terms in the vertices of interactions.   These terms would lead to corrections to  the shear viscosity coefficient proportional to $ (T/T_F)^2$ (or higher powers) and are therefore negligible. In other words, the dispersion law determines the leading temperature dependence of $\eta_{\rm 3ph}$, the vertex corrections lead to sub-leading temperature dependence. 

Given the strong dependence of the shear viscosity coefficient on the phonon dispersion law, one may question whether with increasing temperature higher order terms in the phonon Lagrangian may be relevant. In the Appendix \ref{appendix-disp-law} we argue that the neglected terms should be  relevant only  for temperatures 
$T\sim ( 0.2 -0.3) T_F$; given that  the
 transition to the normal phase takes place at $T_c \simeq 0.16T_F$ \footnote{This is the value of the critical temperature in an homogenous system} \cite{Haussmann:2008, kinast3},  those terms are irrelevant for a discussion of the shear viscosity coefficient in the superfluid phase.

%%%%%%%%%%%%%%%%%%%%%%%%%%%%%%%%%%%
\section{Evaluation of the shear viscosity 
 and comparison with the experiments}
\label{experim-eta}
%%%%%%%%%%%%%%%%%%%%%%%%%%%%%%%%%%%%%%

The experimental measurement of the  shear viscosity of ultracold fermionic atoms is performed confining a degenerate mixture of an equal number of spin-1/2-up and spin-1/2-down in an optical trap. In the following  we shall refer to the experimental setup realized in Refs.~\cite{kinast1, kinast3} with  $N \simeq 2.0 \times 10^5$ $^6$Li atoms  in a magnetic field of strength $B= 840$ G,   close to the $s$-channel Feshbach resonance at $B=834.15$ G \cite{Bartenstein}.   
The unperturbed number density of the trapped fermions is given by 
%(in units  $\hbar = k_B=1$, hereafter used)
\be
\label{trap}
\rho_0({\bf r}) =\frac{(2 m E_F)^{3/2}}{3 \pi^2} \left(1-\sum_{i=1}^3 \frac{r_i^2}{R_i^2}\right)^{3/2} ~~~ R_i=\sqrt{\frac{2 E_F}{m \omega_i^2}} \,,
\ee
where $m$ is the mass of the  $^6$Li atom and $E_F$ is the Fermi energy of $N$ free fermions in a harmonic 
oscillator potential.  In   Refs.~\cite{kinast1, kinast3},  the transverse frequencies of the trap are   $\omega_x\simeq 2\pi\times 1778\,{\rm s}^{-1}$, 
$\omega_y \simeq 2\pi\times 1617\,{\rm s}^{-1}$
and the net axial frequency is $\omega_z\simeq 2\pi\times76\,{\rm s}^{-1}$, corresponding to a  Fermi temperature   $ T_F \simeq 2.4~10^{-6} {\rm K}  $.  Given the asymmetry of the trap,
it is convenient to redefine the radial coordinate as
\be
\omega^2 r^2=\omega_x x^2+ \omega_y y^2+\omega_z z^2
\ee
where $\omega=(\omega_x \omega_y \omega_z)^{1/3}$. Using this definition of the radial coordinate, the system is
spherically symmetric, indeed the trapping potential, defined in Eq.~(\ref{eq:pot}), is given by $V(r)=\frac{1}{2} \omega^2 r^2$. The shear viscosity  can be extracted at both high and low temperatures by studying either the anisotropic expansion of the  atomic cloud or  the breathing mode damping \cite{Cao:2010wa}. 

The critical temperature of the trapped system is $T_c^{{\rm trap}} \simeq (0.21 - 0.25) T_F$ \cite{Haussmann:2008, kinast3}, and corresponds to the temperature for which no region of the system is in the superfluid phase. Whether the system is locally in the superfluid phase depends on the local value of the Fermi temperature, and it is found that the system is superfluid for $T < 0.16 T_F(r)$. Here $T_F(r)$ is the local value of the Fermi temperature in the LDA. The part of the fermionic cloud which is superfluid occupies the trap center, that extends up to a distance $\bar {\bf r} (T)$ such that $T = 0.16 T_F(\bar {\bf r})$.  Since at very low temperature the matter distribution in the trap center
is roughly constant, and the largest part of the fluid is in the trap center, we shall assume that $T_F(r)=T_F$ for $r \leq \bar{\bf r}$. The trap center is surrounded by a corona of normal fluid, which corresponds to the range $r > \bar{\bf r}$, where  $T > 0.16 T_F(r)$. For $T>T_c^{\rm trap}$ one has that $\bar {\bf r}=0$; only fermionic modes are present and their collision processes determine the value of the shear viscosity coefficient.  At high $T$ a simple dimensional analysis, or a much careful study based on a Boltzmann equation, predicts the scaling  $\eta \propto T^{3/2}$ \cite{aboveTc}, in agreement with the experimental results~\cite{Cao:2010wa}. For a recent determination of the shear viscosity using quantum Monte Carlo calculations see Refs.~\cite{Drut,dusling}.

For temperatures  below the superfluid phase transition,  both  the theoretical and experimental studies of the transport properties of the system are quite challenging. In this case the low energy spectrum of the system consists of two types of quasiparticles: the fermions and the phonons. The phonons  are present in the center of the trap, where the superfluid phase is realized, while the fermionic modes are present in the trap center as well as in the corona.  We shall assume that the leading contribution to the shear viscosity is due to the  trap center, where the density is larger.  In the trap center the system is superfluid and therefore  the quasiparticle fermions are gapped, and their density is exponentially suppressed as $e^{-\Delta/T}$, where $\Delta$ is the value of the fermionic gap. Nevertheless, their contribution to the shear viscosity may be sizable at any temperature. Indeed it was shown by Pethick {\it et al.} \cite{Pethick:1975}  that the shear viscosity coefficient of superfluid $^3$He is not exponentially suppressed at low temperature, but tends to a constant value $\sim 1/4 \eta(T_c)$. The case of trapped ultracold fermionic system close to unitarity was studied in a recent publication \cite{Guo:2010dc} and the contribution of fermionic modes to the shear viscosity was analyzed by computing the value of the effective carrier fermion number in the superfluid phase as a function of $T$. It has been concluded that the fermionic modes in the superfluid phase cannot be responsible for the experimental values of the shear viscosity   below $T_c$. Instead, it is suggested that the shear viscosity is dominated by the fermions in the corona of the trap. However, the obtained numerical values of $\eta/s$ are about a factor two larger than the experimentally measured values.

Since  phonons are gapless excitations, their number density is not exponentially suppressed at low $T$, but decreases as $T^3$.  Therefore,  at sufficiently low temperature they should give a sizable contribution to the thermodynamic quantities as well as to the  transport properties  of ultracold fermionic systems. In the BEC limit it is indeed known that below $T_c$, phonons give the dominant contribution to the entropy of the system as well to the shear viscosity coefficient \cite{IntroSupe}. On the other hand, in the BCS limit their contribution is negligible as compared to the fermionic contribution. At unitarity the contributions of fermionic and bosonic modes to the thermodynamic quantities might   be comparable, as shown in  \cite{Diener:2008} studying gaussian fluctuations on the top of a mean field solution. However, it is reasonable to expect that at sufficiently low temperature the contribution of fermions to the entropy of the system will be exponentially suppressed. Therefore, we shall assume that $s = s_{\rm ph}$, given by the expression in Eq.~\eqref{entropy} . 

Regarding the shear viscosity  coefficient, one has to combine the fermionic and the bosonic contributions. One possible and challenging way of doing this, is to solve the Boltzmann equations which take into account both degrees of freedom.   However, since the results of Ref.~\cite{Guo:2010dc}   show that the fermionic contribution to $\eta$ is above the experimental values, there should be a mode with a lower shear viscosity coefficient. Indeed, $\eta$ is directly proportional to the typical time for transport of momentum in the direction orthogonal to the flow, $\tau_{\perp}$,  and therefore a mode with a shorter $\tau_{\perp}$ should exist. We assume that this mode is the phonon. Actually, the authors of  Ref.~\cite{Guo:2010dc}  claim that the interaction between phonons cannot lead to the experimentally measured value of the viscosity, because 
phonons do not couple to transverse probes. They also notice that  the  contribution to  $\eta$ found in Ref.~\cite{Rupak:2007vp} increases  when the temperature is lowered, in disagreement with the experimental results. However, the arguments presented in Ref.~\cite{Guo:2010dc} are not conclusive.
 As we shall show in the next section in the analysis of the low temperature regime, the inclusion of finite size effects in the phonon dynamics leads to a contribution to  $\eta$  that decreases when the temperature decreases.  Furthermore,   there is no physical reason why a quasiparticle of the system that transports energy cannot transport momentum in the direction orthogonal to the direction of the flow. In $^4$He, indeed, it has been shown that at sufficiently low temperature phonons give a sizable contribution to the shear viscosity  \cite{IntroSupe}, and an analogous result should hold for any ultracold superfluid.

When the temperature of the system approaches $T_c^{\rm trap}$ from below, the contribution of fermions to the shear viscosity and to the entropy of the system should increase, for two reasons. In the first place, the radius of the trap center decreases, and  a larger portion of the system is in the normal phase.  Secondly, the value of the quasiparticle gap $\Delta$ should decrease with increasing temperature and therefore close to $T_c^{\rm trap}$ the fermionic modes in the trap center should not be exponentially suppressed. Indeed,  close to $T_c^{\rm trap}$ the thermodynamic description in terms of fermionic modes gives an excellent description of the experimentally measured properties of the system \cite{Haussmann:2007zz, Haussmann:2008}. On the other hand, starting from vanishing temperature and increasing $T$,   both effects mentioned above result in an increasing importance   of the fermionic contributions, however it is not obvious at which temperature the fermionic contribution will dominate the bosonic one. At the present stage, only a comparison with experimental results can help to determine such a temperature. In the following we shall  discuss the contribution of the bosonic modes to $\eta/s$, and show that the very low temperature experimental points can be described by taking into account solely the contribution of phonons.

\subsection{Finite size effects}
\label{finitesection}

As already mentioned in Sec.~\ref{subsec-values}, the contributions to the shear viscosity of 4-ph and 3-ph processes diverge at vanishing temperature. This unphysical behavior is due to the fact we have performed the various calculations in an infinite volume. In order to properly evaluate the contribution of phonons to the shear viscosity, it is important to consider the finite extension of the optical trap  and the corresponding matter distribution. In the LDA the effect of the matter distribution can be taken into account computing the trap average of the shear viscosity evaluated in the previous section.  At very low temperature, the matter
distribution in the trap center is roughly constant, indeed we have assumed $T_F(r)=T_F$ for $r \leq \bar {\bf r}$, and we expect that the trap average would lead to a tiny variation of our results.

However,  the finite extension of the trap does play an important role because the interaction of quasiparticles with the boundary can be neglected only if their mean free path  is sufficiently small.   More precisely, finite size effects have to be taken into account when the Knudsen number $K_n = \frac{l} {L}$, defined as the ratio of the mean free path $l$ of the quasiparticles over the typical size of the system $L$, is not small.

The phonon mean free path $l_{\rm ph}$ can be obtained by considering the same 3-ph and 4-ph processes we have taken into account for the evaluation of the shear viscosity. The shear viscosity and the corresponding mean free path can be readily evaluated by the formula
 \be
\label{bulkvis}
\eta_{\rm bulk} = \frac 15 \rho_{\rm ph}  c_s  l_{\rm ph} \ ,
\ee
where $\rho_{\rm ph} = \frac{2 \pi^2 T^4}{45 c_s^5}$ is the phonon density of the normal fluid component. At low T, for the 3-ph processes one  obtains 
 that $l_{\rm 3ph} \propto 1/T^5 $, while for 4-ph collisions   $l_{\rm 4ph} \propto 1/T^9$.  Thus, no matter which process dominates, for sufficiently low temperature the mean free path becomes larger than the typical length of the system. In the case we are interested in, we take as the typical size the smallest radius of the atomic cloud, $R_x$, given in Eq.~(\ref{trap}).  We present in Fig.\ref{fig:lph} a plot of the value of $l_{\rm ph}/R_x $  as a function of $T/T_F$ for both 3-ph and 4-ph processes. The vertical dashed green line in this plot and in all the subsequent plots indicates the critical temperature $T_c = 0.25 T_c^{\rm trap}$ reported in  \cite{kinast3}. However, the critical temperature might be slightly lower, i.e.  $T_c \simeq 0.21 T_c^{\rm trap}$, as obtained in \cite{Haussmann:2008}. It has to be understood that  the various plots of the phonon mean free paths  are meaningful only for values of $T  < T_c^{\rm trap}$  because phonons do not exist in the normal phase. For the same reason, any plot of the phonon shear viscosity coefficient is meaningful only in the superfluid phase. Regarding the 3-ph processes the mean free path strongly depends on the value of  $\psi_{\rm max}$. As we shall see in the next section, the experimental data seem to favor the values $\psi_{\rm max} \sim 0.2-0.3$, and the lines named 3ph-a, 3ph-b and 3ph-c  correspond to  the mean free  paths  obtained with  $\psi_{\rm max}=0.4$, $\psi_{\rm max}=0.3$ and $\psi_{\rm max}=0.2$ respectively For  $\psi_{\rm max}  \lesssim 0.2$ the mean free path of  the 3-ph process is always larger than $R_x$.  For  $\psi_{\rm max}  > 0.3$ the mean free path of  the 3-ph process becomes of the order of the size of the atomic cloud only for $T \sim 0.12 \,T_F$, and rapidly increases for lower values of $T$.  We also present the plot of  $l_{\rm 4ph}/R_x$, (named 4-ph) corresponding to the dashed black line in Fig.\ref{fig:lph}.  Also in this case the mean free path becomes of the order of the trap radius $R_x$ for $T \sim 0.12 \,T_F$. Notice that the curves 3-ph-a and 4-ph almost overlap for $T \lesssim 0.15 T_F$, however for larger temperatures  4-ph becomes extremely small, while in the 3-ph-a case it remains of the order  of $0.1 R_x$.  Such a difference is due to the fact that the 3-ph shear viscosity has been
evaluated including the term proportional to $\psi (k)$ in the dispersion law of phonons. In this
case the phonon dispersion law has a maximum, which leads to a nonzero minimum of the
corresponding mean free path for a certain value of $T$. The  $\psi(k)$ term was not included in
the evaluation of the 4-ph shear viscosity and for this reason, the 4-ph curve of the mean
free path does not have a minimum and should not be very reliable when getting close to
$T_c^{\rm trap}$ . As a check we have evaluated the mean free path for the 3-ph processes considering $\psi_{\rm max}  \gg 1 $ (corresponding to $\delta \ll \gamma^2$, see Eq.\eqref{gamma}), meaning that the effect of the maximum
in the phonon dispersion law manifests at very high momenta (and temperature). In the
temperature range reported in Fig. \ref{fig:lph} we have obtained a plot of the mean free path in
concordance with the results that obtained for $\delta=0$ and very similar to the one of the 4-ph
processes. 
 
%Such a difference is due to the fact that in the evaluation of the 3-ph shear viscosity we have included in the dispersion law of phonons the term proportional to $\psi(k)$, while such a term was not included in the evaluation of the 4-ph shear viscosity. For this reason,  the 4-ph curve of the mean free path should not be very reliable when getting close to $T_c$.
% When evaluating the 3-ph shear viscosity  we have considered the expression in Eq.\eqref{psik}, which has a maximum for a certain value of $k$ (see Appendix \ref{appendix-disp-law}),  parameterized by $\psi_{\rm max}$. When the effect of the maximum is neglected, i.e.  $\psi_{\rm max}  \gg 1 $ (corresponding to $\delta \ll \gamma^2$, see Eq.\eqref{gamma}),  one obtains for the 3-ph processes a plot of the mean free path very similar to the one of the 4-ph processes, in concordance with the results that are obtained when $\delta=0$.

\begin{figure}[ht]
\epsfxsize=12truecm
\centerline{\epsffile{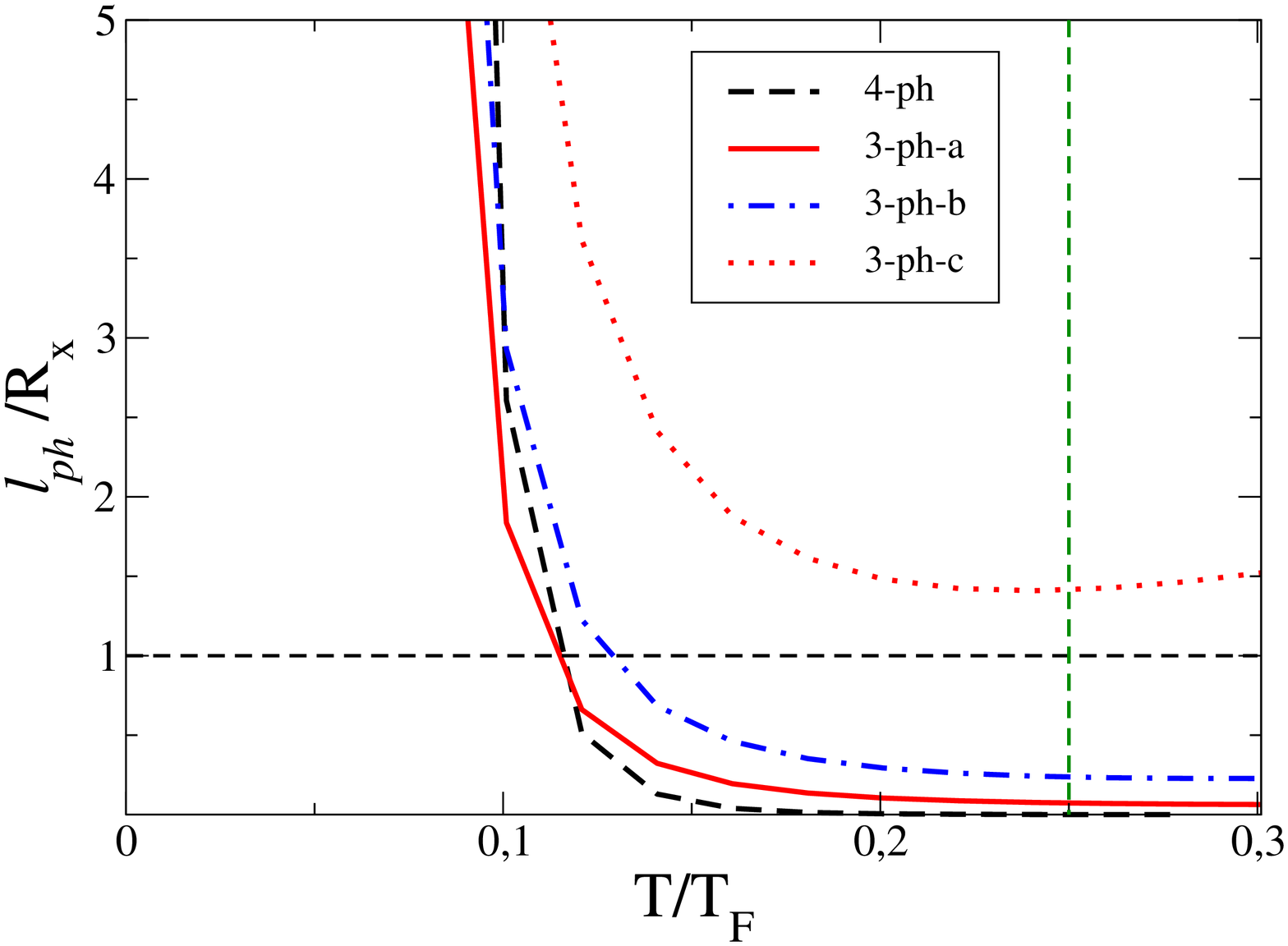}} \noindent
\caption{\it Plot of  the Knudsen number, $l_{\rm ph}/R_x$,  as a function of $T/T_F$ for 3-phonon (3-ph) and 4-phonon (4-ph) processes. The black dashed line corresponds to the mean free path associated to the 4-ph process. The solid red line (named 3-ph-a) corresponds to the  3-ph process with $\psi_{\rm max}=0.4$, the  dot-dashed blue line (named 3-ph-b) corresponds to the 3-ph process with $\psi_{\rm max}=0.3$ and the dotted red line (named 3-ph-c) corresponds to the  3-ph process with $\psi_{\rm max}=0.2$.  The horizontal black dashed line corresponds to  $l_{\rm ph}/R_x=1$. The hydrodynamic description is valid for $l_{\rm ph}/R_x \ll 1$. The vertical dashed green  line approximately corresponds to the transition temperature between the normal phase and the superfluid phase \cite{kinast3}.
\label{fig:lph}}
\end{figure}

When the mean free path of the phonons  exceeds  the typical size of the superfluid region, the transport properties of the phonons are mainly governed by their interactions with the boundary, rather than by the self-interactions. In the trapped atomic system the density is not constant, and the superfluid region at the trap center is surrounded by  a corona  of a dilute fluid  of  unpaired fermions. This layer effectively acts as a  boundary for the superfluid phonons, which are confined to the superfluid center of the cloud, and  when
phonons reach this region they are absorbed or scattered back. If the process is elastic, then the phonon will only exert a normal pressure on the boundary, resulting in the propagation of low energy excitations in the normal phase corresponding to standard first sound. But, if the phonon is diffused, then it will as well exert a shear stress on the boundary inasmuch as  the particles of a rarefied gas produce a shear stress when diffused by the walls of a container. 

In principle, both kind of processes, self-interactions and interactions with the boundary, take place for any value of the Knudsen number. However in the hydrodynamic regime, for $K_n \ll 1$, self-interactions are the dominant microscopic processes leading to the largest contribution to the transport  coefficients, while the interactions with the boundary are only relevant in a layer of length $l_{\rm ph}$ (the Knudsen layer) close to the boundary.  The interaction with the boundary can be  effectively taken into account in the hydrodynamic equations by requiring that the fluid velocity  vanishes at the boundary  \cite{Cercignani, Sone, Struchtrup}, the so-called ``no-slip" boundary condition.

Finite size effects in the flow of rarefied gases and fluids have been studied thoroughly in the past  \cite{Cercignani, Sone, Struchtrup}. Depending on the value of $K_n$ they can be taken care of either by imposing a ``slip" boundary condition in  the solution of the Navier-Stokes equations, or, at a microscopic level,  by using the Boltzmann equation with appropriate boundary conditions.
For values of $K_n \lesssim 0.1$  (say) the Navier-Stokes equation are still valid, but the shear viscosity coefficient (and the transport coefficients in general) must be corrected including the finite size corrections, which for low Knudsen numbers can be expressed as 
\be
\label{effective- eta}
\frac{1}{\eta}_{\rm slip} = \frac{1}{\eta_{\rm bulk}} \left( 1 + \frac{\zeta l}{L}\right)\ \ ,
\ee
where $\eta_{\rm bulk}$ refers to the value of the viscosity in the hydrodynamic regime,  and $\zeta$ is a coefficient that characterizes the interaction of the quasiparticles with the boundaries for a particular geometry of the system.
Eq.~(\ref{effective- eta}) gives account to a correction to the standard hydrodynamic equations proportional to $K_n$. 
With increasing $l$, terms of higher order in $K_n$ must be added, but the computation of such terms is complicated even for an ideal gas \cite{Cercignani, Sone}. In any case, with increasing values of $K_n$ one is forced to abandon the Navier-Stokes equation and  more moments of the distribution function have to be considered.   For recent progress  with Grad's moment method  see e.g. \cite{Struchtrup}.

For  values of $K_n  >> 1$,  the system is in the ballistic (also named Knudsen) regime and no hydrodynamic description can be employed for the description of the phonon gas. In this case the interactions of phonons with the boundary are more frequent than the interactions among phonons, and  the relevant relaxation time  is the time between two collisions of a particle with the boundary, $\tau_b$.   As an admittedly rough model of the interaction between the  phonon fluid and the fermions in the external corona, we assume that the boundary between the superfluid and the normal region is a sharp surface, i.e. a {\it simple boundary} \cite{Sone}, and, as in the Maxwell model \cite{Sone, Struchtrup}, the scattering  at the boundary is described by one single parameter, the {\it accommodation coefficient} $\chi$, meaning that particles  impinging on the boundary are diffused with  probability $\chi$ and specular-reflected  with probability $1-\chi$. This simplified description allows us to qualitatively capture the physical process happening at the interface, where a shear stress is exerted by  the phonon gas  on the fermionic fluid of  the external corona. A more detailed and realistic  modeling of  the  interaction of phonons  with the normal fermions is certainly possible, however such a description would require the knowledge of the matter distribution, as well as the computation of the interaction of phonons with (almost) gapless fermionic modes. Moreover, for a proper description of the interface between the normal and superfluid component one should take into account  the transfer of matter between the two phases. This effect can be taken into account by introducing  one  further  phenomenological parameter, usually named the {\it condensation coefficient}  \cite{Sone}, which is related to matter transfer between the two phases.  We postpone a detailed  analysis of the interface to future work and in the present paper we content ourselves with a phenomenological formula for the shear viscosity coefficient which has been used, derived and tested for the description of finite size effects in the flow of rarefied gases and fluids, see e.g. \cite{Hojgaard}, in the presence of a simple boundary.

As in a rarefied gas,  the phonons will exert on the boundary a shear stress which is proportional to the energy density of phonons and will damp the oscillation of the boundary. The same phenomenon takes place in $^4$He at extremely low temperature, $T \lesssim 0.5$ K, where it has been observed that ballistic  phonons  can efficiently damp the movement of immersed objects. Experiments done with an oscillating sphere \cite{Eselson}, a vibrating microsphere \cite{Niemetz}, or  a vibrating quartz tuning fork \cite{Zadorozhko} immersed in superfluid $^4$He  (see \cite{Zadorozhko} also for a comparison of the results obtained with various experimental apparatuses) show that the damping of  the oscillations can be described, in perfect accommodation  ($\chi =1$), by the introduction of a ballistic (effective) shear viscosity coefficient defined by 
\be\label{etaball-He4}
\eta_{\rm ball}  \equiv \frac{1}{5}  \rho_{\rm ph} c_s  d\,,
\ee
where $d$ is the typical size of the oscillating object. This expression leads to excellent agreement with the experimental data for large Knudsen numbers \cite{Zadorozhko}, predicting the correct dependence on the temperature and on  the  typical size of the oscillating object.

Note that the expression above is equivalent to the one in Eq.\eqref{bulkvis} with the replacement of $l_{\rm ph}$ with $d$ and suggests that in the ballistic regime  (and in the presence of a boundary) one can extend the standard definition of the transport coefficient by replacing the mean free path with the typical size of the system. Therefore, in our case, we define a ballistic (effective) shear viscosity as
\be\label{ball}
\eta_{\rm ball} = \frac{1}{5}  \rho_{\rm ph} c_s \chi (c_s \tau_b) \equiv \frac{1}{5}  \rho_{\rm ph} c_s  a\,,
\ee
where $c_s  \tau_b= L$ is  the typical size of the system, i.e. the extension of the superfluid region at the trap center (which is smaller than $R_x$).  In the present paper we are not interested in a detailed description of the interface between the normal and the superfluid phase but rather to understand whether phonons can give a sizable contribution to the shear viscosity coefficient. For this reason,  we define the phenomenological length  $a=\chi L$, which we will use as a fitting parameter.

In order to evaluate the shear viscosity in the intermediate region, $K_n \sim 1$, one should employ the Boltzmann equation treating collisions of phonons with the boundary and collisions among phonons on an equal footing.  Solving such a problem is complicated even for ideal gases \cite{Cercignani, Sone, Struchtrup}. However, since we know the behavior of the relaxation time for large values of $K_n$ as well as for $K_n \ll 1$,  it is  possible to employ the same reasoning at the basis of the Matthiessen's rule \cite{Ashcroft-Mermin}  to obtain an expression of $\eta$   in the intermediate region, which  interpolates between the values of the shear viscosity coefficient in the Knudsen and in the hydrodynamic regimes. For this purpose we define an effective relaxation time, $\tau$, incorporating the effects of inter-particle collisions and of the collisions with the boundary  
\be \tau^{-1} = \tau^{-1}_b + \tau^{-1}_{\rm ph}
\,.\ee

This relation follows from the assumption that the total collision frequency is the sum of the frequencies of the two mentioned kinds of collisions, thus we are assuming  that the two collision processes are not correlated. Since the shear viscosity coefficient is proportional to the collision time,  we define the total effective shear viscosity as
\be
%\label{eff-eta}
\eta_{\rm eff} = \left( \eta_{\rm  3ph}^{-1} + \eta_{\rm ball}^{-1}\right)^{-1} \,,
\label{eff}
\ee
where $\eta_{\rm ball}$ is defined in Eq.\eqref{ball} and $\eta_{\rm 3ph}$ is the 3ph shear viscosity. 
In principle, the contribution of the 4ph collisions should be considered as well.
However, the $\eta_{\rm 4ph}$ has been computed assuming $\psi(k)=0$. This approximation should not be
of great impact  in the evaluation of the shear viscosity  at low temperature as  shown in Refs.~\cite{Rupak:2007vp,Manuel:2011ed}. However, it affects the higher temperature behavior of   $\eta_{\rm 4ph}$, and for this reason it cannot be included consistently in our analysis. That is why we will neglect binary collisions in the following subsection.

\subsection{Comparison with the experimental data}\label{sec-exp}

The experimental values of the shear to entropy ratio have been obtained  by the Duke group in a number of different settings. In the present paper we refer to the data set reported in Ref.~\cite{Cao:2010wa} and we show that superfluid phonons can explain the low $T$ experimental behavior of  the shear viscosity coefficient. The caveat is that  in our analysis we use several simplifying assumptions and two fitting parameters. Let us summarize our assumptions. We consider  that the only contributions to the shear viscosity are due to processes taking place among phonons  or between phonons and fermions at the interface between the superfluid and the normal phase. We neglect, instead, contributions to $\eta$ due to interactions between the fermions. 
In other words we assume  that any   oscillation of the corona is damped by the friction with the trap center, and this process is effectively described by the ballistic shear viscosity in Eq.~\eqref{ball}. With increasing temperature, the contribution to $\eta$ of the interaction processes among fermions should become more and more relevant due to the fact that the spatial extension and the density of the external corona increases and because more Bogolyubov modes are available. Therefore our results should be reliable only up to some temperature $T < T_c$. 

Regarding the phonon contribution, we also neglect corrections to the shear viscosity coefficient coming from the higher energy part of the phonon spectrum. This amounts to neglect terms of order $(k/k_F)^7$ in the phonon dispersion law. In the Appendix \ref{appendix-disp-law} we show that these terms should be negligible for temperatures up to $(0.2-0.3) T_F$. 

While our formalism allows us to compute  the shear viscosity in any point of the trap, performing the trap average is not a simple task.  Thus, we assume that the largest contribution comes from the trap center, where the density is larger and almost constant, and where
the local Fermi temperature, $T_F (r)$, is roughly constant and we approximate it with the
global value $T_F$.

In order to explain the low $T$ data we will use Eq.~(\ref{eff}), which depends on two unknown parameters, $a$ and $\delta$ (or equivalently $\psi_{\rm max}$). We remind that the value of $\delta$ could be in principle determined if the phonon dispersion law is computed to next to next to leading order,
 while  $a$ requires to know both the size of the superfluid core in the trap and the probability of the phonons to be  diffused at the edges of the core.  

In Fig.~\ref{fig:eta-s-3psi} the  experimental values of $\eta/s$ obtained in \cite{Cao:2010wa} are shown, together with three fits obtained by considering the contribution of the interaction of phonons  with the boundary, described by Eq.\eqref{ball}, assuming $a=0.3\,R_x$,  and the 3-ph  processes with $\psi_{\rm max} =0.2$ (dashed black line),  $\psi_{\rm max} =0.3$ (dotted red line) and  $\psi_{\rm max} =0.4$ (solid blue line).  The experimental data seem to favor the value $\psi_{\rm max} =0.3$.  However, more experimental data  are needed in order to figure out the correct value of $\psi_{\rm max}$.  Note also that the experimental values of $\eta/s$ decreases with decreasing temperature and this behavior is well reproduced for $\psi_{\rm max} \lesssim 0.3$. Although all the experimental values are above the value $1/4\pi$ it seems that further decreasing the temperature values below the universal bound might be reached. 

In Fig.~\ref{fig:eta-s-3a} we present three fits of the experimental data considering the contribution of the 3-ph  processes (with $\psi_{\rm max} =0.3$) and the interaction of phonons with the boundary, effectively describe by  Eq.\eqref{ball}  with  $a=0.2\,R_x$ (dashed black line), $a=0.3\,R_x$  (dotted  red line)  and  $a=0.4\,R_x$ (solid blue line). The experimental data seem to favor $a=0.3\,R_x$, but  more experimental data  are needed to determine the precise value of this parameter. 

\begin{figure}[ht]
\epsfxsize=12truecm
\centerline{\epsffile{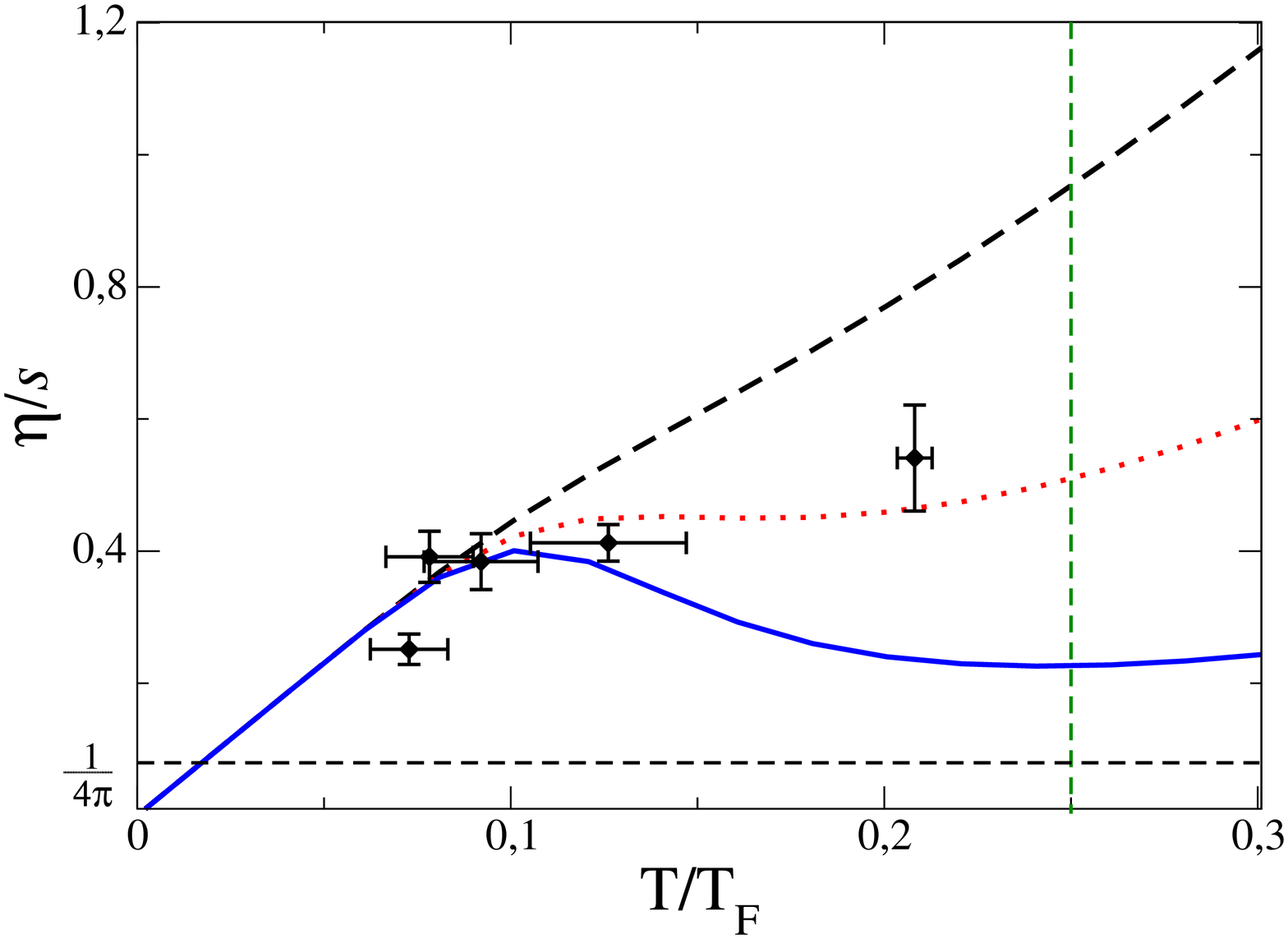}} \noindent
\caption{\it  Plot of  $\eta_{\rm}/s$ in units of  $\hbar/k_B$ as a function of $T/T_F$ considering the contribution of the interaction with the boundary, assuming $a=0.3\,R_x$,  and the 3-ph  processes with three different values of $\psi_{\rm max}$. The dashed black line is obtained  with $\psi_{\rm max} =0.2$, the dotted red line   is obtained  with $\psi_{\rm max} =0.3$ and the solid blue line is obtained with $\psi_{\rm max} =0.4$.    The experimental values and the corresponding error bars were taken from \cite{Cao:2010wa}. The vertical dashed green  line approximately corresponds to the transition temperature between the normal phase and the superfluid phase in a trap reported \cite{kinast3}. The actual value of $T_c^{\rm trap}$ might be slightly lower; the value reported in \cite{Haussmann:2008} is $T_c^{\rm trap} \simeq 0.21 T_F$.
\label{fig:eta-s-3psi}}
\end{figure}

\begin{figure}[ht]
\epsfxsize=12truecm
\centerline{\epsffile{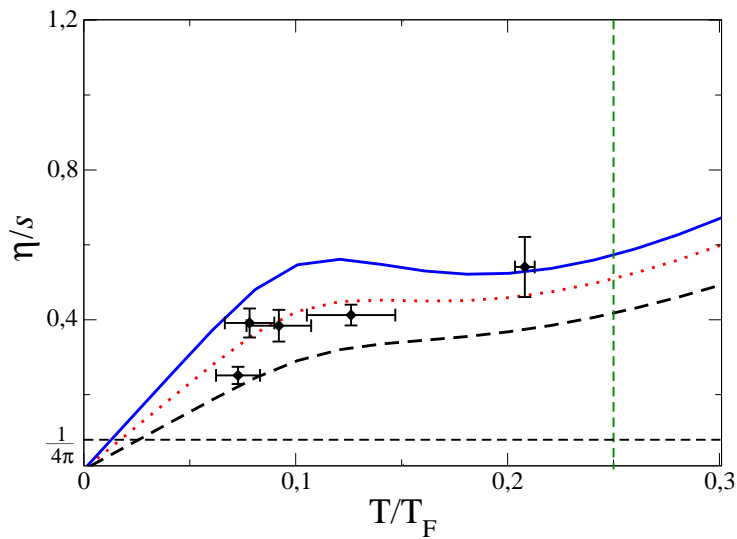}} \noindent
\caption{\it Plot of  $\eta/s$ in units of  $\hbar/k_B$ as a function of $T/T_F$ considering the contribution of the interaction with the boundary and the 3-ph  processes with $\psi_{\rm max} =0.3$, and  $a=0.2\,R_x$ (dashed black line), $a=0.3\,R_x$  (dotted  red line)  and  $a=0.4\,R_x$ (solid blue line).   The experimental values were taken from \cite{Cao:2010wa}. The vertical dashed green  line approximately corresponds to the transition temperature between the normal phase and the superfluid phase \cite{kinast3}. 
\label{fig:eta-s-3a}}
\end{figure}

In Fig.\ref{fig:eta-s-total} we report several plots of all the various contributions separately as well as the effective  shear viscosity. The interaction of the phonons with the boundary (dashed black line)  is obtained from Eq.~\eqref{ball} assuming $a=0.3\,R_x$; the contribution of the 3-ph  process (dotted  red line) is obtained from Eq.~\eqref{eta-3phallT} considering  $\psi_{\rm max} =0.3$; the contribution of the 4-ph process  (dot dashed green line) is obtained from Eq.~\eqref{4pheta}. The effective shear viscosity (solid blue line) is obtained combining $\eta_{\rm ball}$ and  $\eta_{\rm 3 ph}$ in  Eq.~\eqref{eff} with the above reported values of $a$ and $\psi_{\rm max}$. With this choice of the parameters the experimental values are appropriately reproduced.

We refrain from showing results for the effective shear viscosity including 4ph processes. Actually, for sufficiently low temperature the expression of $\eta_{\rm 4ph}$  reported in Eq.~\eqref{4pheta} is reliable and it can be included in the computation of the effective shear viscosity; but in this case the contribution of 4ph processes is however negligible, because of the dominant role of the ballistic viscosity. For higher temperature  Eq.~\eqref{4pheta} is not reliable and NLO contributions of  the phonon dispersion law should be included.

Our plots show  that at very low temperatures,  below $0.1 \, T_F$, the ballistic shear viscosity dominates, meaning that the only relevant dissipative mechanism is the one that takes place at the interface between the normal and the superfluid phase. This result is independent of the detailed expression of the phonon dispersion law, in particular this result remains correct  also if 3-ph processes are not allowed, i.e. considering  $\gamma <0$, and only  binary collisions are included  in the analysis.  It only depends on the fact that at sufficiently low temperature phonons become ballistic, see Fig. \ref{fig:lph}.

\begin{figure}[ht]
\epsfxsize=12truecm
\centerline{\epsffile{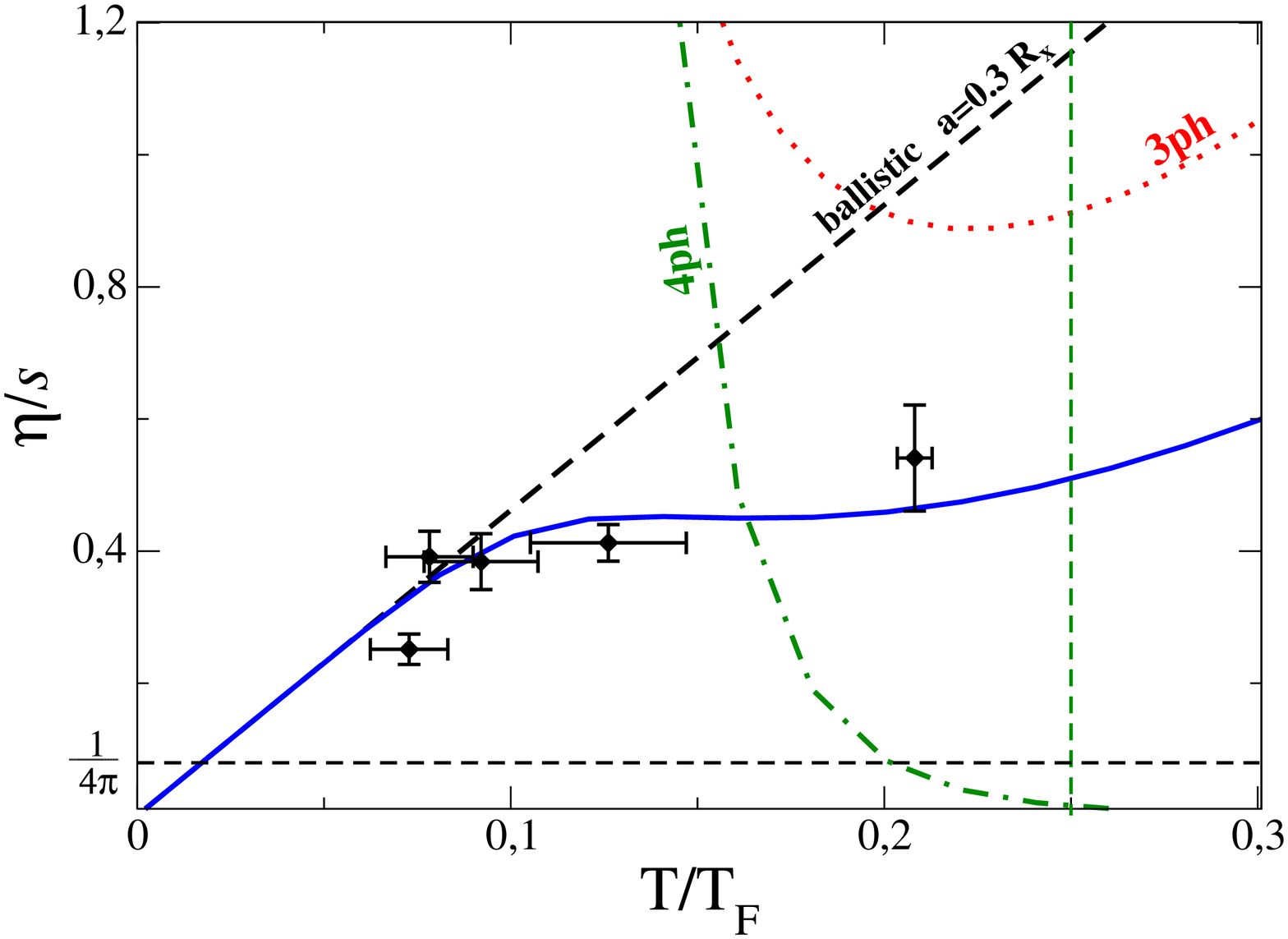}} \noindent
\caption{\it  Plot of  $\eta/s$ in units of  $\hbar/k_B$ as a function of $T/T_F$. We report all the various contributions separately as well as the effective  shear viscosity obtained with Eq.\eqref{eff}.  The interaction of the phonons with the boundary (dashed black line)  is obtained from Eq.\eqref{ball} assuming $a=0.3\,R_x$; the contribution of the 3-ph  process (dotted  red line) is obtained from Eq.\eqref{eta-3phallT} considering  $\psi_{\rm max} =0.3$; the contribution of the 4-ph process  (dot dashed green line) is obtained from Eq.\eqref{4pheta}.    The experimental values and error bars were taken from \cite{Cao:2010wa}. 
The solid blue line is obtained with Eq.~\eqref{eff} combining the ballistic term and 3-ph processes. 
The vertical dashed green  line approximately corresponds to the transition temperature between the normal phase and the superfluid phase \cite{kinast3}. The actual critical temperature may be lower, $T_c^{\rm trap} \simeq 0.21 T_F$ as reported in \cite{Haussmann:2008}.}
\label{fig:eta-s-total}
\end{figure}

\section{Summary}

Let us give a very brief summary of our results. Below the critical temperature the trapped Fermi gas  is in a superfluid phase, characterized by the spontaneous breaking of a $U(1)$ symmetry. In this regime there are two types of quasiparticles: the fermions and the  phonon (Nambu-Goldstone) modes. Phonons are gapless modes and at sufficiently low $T$ they become the only relevant degrees of freedom, therefore there exists a temperature where  gapped fermionic modes can be neglected because thermally suppressed. The precise value of this temperature is presently unknown,  however  experimental data \cite{Ku} and Quantum Monte Carlo simulations \cite{Bulgac:2008zz} do not exclude that already at $T\simeq 0.1 T_F$ phonons give a sizable or even leading contribution to the thermodynamic and transport properties of the system.
At unitarity, EFT techniques and  numerical methods can be used to write down the phonon Lagrangian and to assess  the main phonon self-interactions, but  this information it still incomplete, because not all the terms in the Lagrangian beyond the NLO have been determined, so far.  Further and complementary information can  be deduced by the analysis of the experimental data. This mixed approach, already used for different superfluids such as  $^4$He,  allows us to  constrain some parameters of the effective low energy Lagrangian and to understand the most important physical processes.

We have presented a detailed computation of the contribution of small-angle collisions to the shear viscosity.  The $T$ dependence of such a contribution strongly depends on the form of the phonon dispersion law, as it had previously been discovered in $^4$He \cite{Maris}. In this regard,  one notices that the knowledge of the value of $\eta$  in the $T$ regime where it is dominated by phonons would certainly provide detailed information on their dispersion law.

We have then discussed the relevance of the restricted geometry  for  trapped Fermi gases in the evaluation of the shear viscosity. Below $T_c^{\rm trap}$ the superfluid phase is realized in the core of the gas cloud, while being surrounded by an outer layer of fermions in the normal phase. At $T \simeq 0.12 T_F$, the phonon mean free path becomes of the same order, or even larger, than the size of the cloud, and phonons are then in the ballistic regime. Then, phonons  collide more often with the boundary of the superfluid
core, that is, with the outer layer of normal fermions,  than among themselves. Such collisions can provide a shear stress on the boundary, which then results in the damping of  the breathing modes. This dissipative mechanism can still be effectively described by a ballistic shear viscosity. The same phenomena  occur in $^4$He, where it has been observed that ballistic phonons can damp the movement of immersed objects in the superfluid.

In order to describe the measured values of $\eta$ for temperatures in  between the ballistic regime and the hydrodynamic regime, we have used a phenomenological formula for $\eta$, given in Eq.~(\ref{eff}), which contains two (yet) unknown parameters, that we fit to the data. One parameter describes the phonon dispersion law at next-to-next-to-leading order, which in principle could be computed from the microscopic physics, while the other one is related to the size of the superfluid core in the trap and the type of phonon-fermion scattering processes that take place at the boundary. 
 
%As already discussed in the previous section,  we have used  a number of simplifying assumptions which we briefly summarize here. We have assumed that the shear viscosity is dominated by the center of the trap, where the density is bigger and roughly constant, which assures that  performing a  trap average might not modify drastically our results.
%We have neglected the fermionic contribution to $\eta$, both in the superfluid core and in the outer normal layer. This should be a valid approximation at low $T$, where superfluid fermions are known to be exponentially suppressed, while the outer fermionic layer of the cloud might be too dilute to provide enough damping. Finally, we have modeled the superfluid-normal fluid interface as a sharp boundary, where phonons are specular-reflected or diffused according with the Maxwell model \cite{Cercignani, Sone, Struchtrup}. We leave for future work  a  quantitative estimate of how good all the above approximations are. 

%A couple of testable predictions can be made from our study. 

Although our model relies on a number of simplifying assumption, detailed in the
previous section, and, as a consequence, gives a rough description of the experimental
system, it nevertheless allows us to make a couple of testable semi-quantitative predictions.
First, we notice that if experiments are conducted reducing/increasing the size of the trap, but keeping $E_F$ constant, then $\eta/s$  at low temperatures should decrease/increase. In other words, the value of $\eta/s$ should correlate with the size of the
gas cloud. Second, we predict that $\eta/s$  should decrease with decreasing temperature. If we naively
extrapolate our results to lower temperature, we predict that there should be a
violation of the string theory proposed bound of $\eta/s$. Note, however, that this violation happens because phonons are  ballistic, while the string theory bound concerns the  hydrodynamic regime. Both predictions are independent of the detailed form of the phonon dispersion law, in particular, they are independent of the sign of the $\gamma$ term in Eq~\eqref{NLOdisp-law}. However, if $\gamma >0$, meaning that in a certain range of momenta the Beliaev processes are allowed,  it is possible to deduce further information on the phonon dispersion law from the experimental data. In particular, we find that the correction to the dispersion law should be given by the expression reported in Eqs.~\eqref{disp-law} an \eqref{psik} with $\delta \simeq 0.03/k_F^4$.

\acknowledgements{We thank J.E.~Thomas and C.~Cao for providing us with the experimental data points of the shear to entropy ratio. 
M.M. thanks W. Zwerger for discussion and suggestions. This research was supported in part by Ministerio de Ciencia e Innovaci\'on under contract FPA2010-16963. LT acknowledges support from the Ramon y Cajal Research Programme from Ministerio de Econom\'{\i}a y  Competitividad and from FP7-PEOPLE-2011-CIG under Contract No. PCIG09-GA-
2011-291679.}

\appendix
\section{Phonon dispersion law and possible ``roton'' excitations}\label{appendix-disp-law}

From the fit of the experimental values of the shear viscosity  performed in Sec. \ref{sec-exp} we have determined the coefficient $\delta$ of the phonon dispersion law in Eq.\eqref{psik}.
 In order to emphasize the region of validity of the expansion of the phonon dispersion law obtained, we rewrite it as an expansion in $x=k/k_F$,
\be\label{phonon-dl}
E/c_s k_F =  x \left(1 + \tilde \gamma x^2 - \frac{\tilde \gamma^2}{4 \psi_{\rm max}} x^4\right)\,,
\ee
where we have used Eq.\eqref{gamma}, and we have defined $\gamma=\tilde\gamma/k_F^2$. The corresponding plot, taking $\psi_{\rm max}=0.1$ (solid black line),  $\psi_{\rm max}=0.2$ (dashed blue line), $\psi_{\rm max}=0.3$ (dotted red line) is reported in Fig. \ref{fig:phonon-dispersion} for positive values of $E$. 
\begin{figure}[ht!]
\epsfxsize=12truecm
\centerline{\epsffile{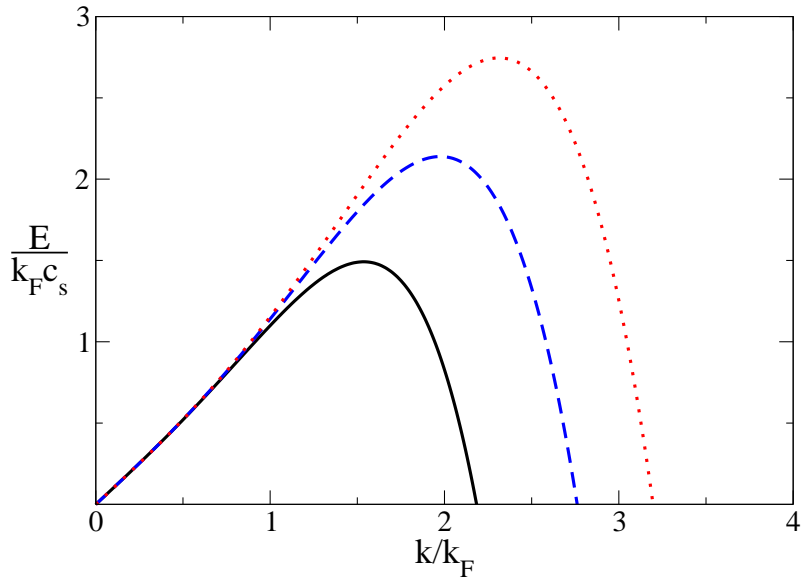}} \noindent
\caption{\it  Plot of  the dispersion law of phonons  given in Eq.~\eqref{phonon-dl}, with $\psi_{\rm max}=0.1$ (solid black line),  $\psi_{\rm max}=0.2$ (dashed blue line), $\psi_{\rm max}=0.3$ (dotted red line).}
\label{fig:phonon-dispersion}
\end{figure}

This plot allows us to figure out whether the neglected terms in the phonon dispersion law are relevant for the present analysis of the shear viscosity coefficient. The terms not included in the momentum expansion are of the order of $x^7=(k/k_F)^7$ and one would naively expect that they should give a large contribution to the dispersion law, and to the integral in Eq.\eqref{theMinteg}, as soon as $k \sim k_F$. However,  the expansion in Eq.\eqref{phonon-dl}  suggests that the actual expansion parameter is not $x$, but $\sqrt{\tilde \gamma} x$ and therefore the naive power counting in powers of $k/k_F$ is incorrect. Assuming that the expansion is in powers of  $\sqrt{\tilde \gamma} x$ and considering that $\tilde \gamma \simeq 0.18$, see Eq.\eqref{NLOdisp-law}, it follows that  Eq. \eqref{phonon-dl} gives a good approximation as far as  $k \lesssim 2 k_F$. This result is corroborated by the observation that the neglected terms should lead to an increase of the phonon dispersion law at $k \sim (2-3) k_F$ in order to avoid that the dispersion law becomes negative, but should not be relevant at smaller values of $k$ (otherwise it would lead to the wrong behavior of the shear viscosity at low temperature). Therefore a term, presumably of order $\tilde\gamma^3 (k/k_F)^7$ should exist, with positive sign, and lead to a minimum of the dispersion law; such a minimum should correspond to rotons. Regarding the temperature range at which such a term would be relevant, we can estimate it considering that the  largest contribution to the shear viscosity coefficient comes from momenta $k \simeq 10\, T/c_s$. Thus the present analysis is consistent up to temperatures $T\simeq (0.2 -0.3) \,T_F$, that is for temperatures of the order or larger than the transition temperature. Thus,  the neglected part of the spectrum  should be important only  close to $T_c$. 

We note that the presence of a dip in the phonon spectrum, both in the BCS and BEC regimes, has been excluded by the study of gaussian quantum fluctuations on the top of a mean field solution  \cite{Diener:2008}. However, the procedure employed in this calculation is only reliable in the BCS and in the BEC limits and it is not clear how a self-consistent calculation might be done close to unitarity (see Section IX of \cite{Diener:2008}). Moreover, higher order fluctuations of the radial oscillation  of the condensate (not included in that analysis) might be relevant close to unitarity, possibly affecting the phonon dispersion law.

\section{Variational treatment with orthogonal polynomials}
\label{ort-poly}

In this Appendix we show that with a different choice of family of variational functions
we reach to the same value of the shear viscosity as that one found in Sec.~\ref{hydro}.
Following a treatment very similar to the one carried out in Ref.~\cite{Rupak:2007vp}  we select the trial functions as
\be
\label{trialfunction}
 h(p)= p^n \sum_{s=0}^{\infty} b_s B_s(p) 
 \ ,\ee
 where $n$ is a parameter that will be determined by a variational procedure, and $B_s(p)$ are 
  orthogonal polynomials of order $s$ and are defined such that the coefficient of the
  highest power $p^s$ is 1, and the orthogonality condition
  \be
  \int \frac{d^3p}{(2\pi)^3} \frac{   f_{\rm eq}(1+f_{\rm eq}) }{2 E_p} p_{ij} p_{ij} p^n B_r(p) B_s(p) =A_{r}^{(n)} \delta_{rs}
    \ee 
 is satisfied.
 
 Using this form of the solution one can check that Eq.~(\ref{1stexpsh}) gives
 \be
 \label{shear1}
 \eta_{\rm 3ph}= \frac{2 c_s^2}{5T} b_0 A_0^{(n)}  \ , \ee 
  where
 \be
 A_0^{(n)} = \frac{T^{6+n}}{6 \pi^2 c_s^{7+n}} \Gamma(6+n) \zeta(5+n)  \  ,
   \ee
 where $\Gamma(z)$ and $\zeta(z)$ stand for  the Gamma and Riemann zeta functions, respectively.
 
On the other hand, the expression of the shear viscosity given by Eq.~(\ref{2onexpsh}) reads
 \be
 \label{shear2}
 \eta_{\rm 3ph} = \sum_{s,t=0}^{N=\infty} b_s b_t M_{st}  \ ,
  \ee
 where
 \be
 M_{st} =   \frac {1}{5T}   \int_{p,p',k'} 
  (2\pi)^4 \delta^{(4)}( P-P'-K')  |{\cal M}|^2 
 f^{\rm eq}_p  ( 1+f^{\rm eq}_{p'})(1+  f^{\rm eq}_{k'}) \Delta_{ij}^s \Delta_{ij}^t  \ ,
  \label{eq:mst}
  \ee
 and
 \be
 \Delta_{ij}^t = B_t(p) p^n p_{ij}  - B_t(p') p'^n p'_{ij} - B_t(k') k'^n k'_{ij}  \ ,
    \ee

Requiring that the two forms of the shear viscosity,  Eqs.~(\ref{shear1}) and (\ref{shear2}), to be equal implies
\be
b_0 = \frac{2 c_s^2}{5T} A_0^{(n)} ( M^{-1})_{00}  \ ,
\ee
so that
\be
\eta_{\rm 3ph} = \frac{4 c_s^4}{25 T^2} (A_0^{(n)})^2 ( M^{-1})_{00} \ .
\ee

In practical terms one performs the study by limiting the number of orthogonal polynomials included
in the analysis. One can prove that 
\be
\eta_{\rm 3ph} \geq  \frac{4 c_s^4}{25 T^2} (A_0^{(n)})^2 ( M^{-1})_{00}  \ ,
\ee
for a particular value of $n$ and the number of orthogonal polynomials considered in the study.

For definiteness, we consider $\psi(p) = \gamma p^2$. We then find that $n=-1$ leads to the
maximum value of $\eta_{\rm 3ph}$. Further, after using only up to the polynomial of order $s=2$ we get the
convergence  to an expression  which matches that of Eq.~(\ref{etagamma}).


\begin{thebibliography}{99}



%--------------------------review------------------------
%\cite{Schafer:2009dj}
\bibitem{Schafer:2009dj} 
  T.~Schafer and D.~Teaney,
  %``Nearly Perfect Fluidity: From Cold Atomic Gases to Hot Quark Gluon Plasmas,''
  Rept.\ Prog.\ Phys.\  {\bf 72}, 126001 (2009)
  [arXiv:0904.3107 [hep-ph]].
  %%CITATION = ARXIV:0904.3107;%%
  
\bibitem{Adams}  
  A. Adams , L.D. Carr, T. Schaefer, P. Steinberg, J.E. Thomas, 
``Strongly correlated quantum fluids: ultracold quantum gases, quantum chromodynamic plasmas, and holography duality",
[arXiv:1205.5180 [hep-ph]]- 
  
\bibitem{Giorgini:2008zz}
  S.~Giorgini, L.~P.~Pitaevskii and S.~Stringari,
  %``Theory of ultracold atomic Fermi gases,''
  Rev.\ Mod.\ Phys.\  {\bf 80}, 1215 (2008).
  %%CITATION = RMPHA,80,1215;%%

%--------------------------HO------------------------

%\cite{Ho:2004zza}
\bibitem{Ho:2004zza} 
  T.~-L.~Ho,
  %``Universal Thermodynamics of Degenerate Quantum Gases in the Unitarity Limit,''
  Phys.\ Rev.\ Lett.\  {\bf 92}, 090402 (2004)
  [cond-mat/0309109].
  %%CITATION = COND-MAT/0309109;%%





  



%------------Experiments-------------------
\bibitem{Ohara:2002}
K.~M.~O'Hara, S.~L.~Hemmer, M.~E.~Gehm, S.~R.~Granade and J.~E.~Thomas, Science {\bf 298}, 2179
(2002) [arXiv:cond-mat/0212463];  M.~E.~Gehm, S.~L.~Hemmer, S.~R.~Granade, K.~M.~O'Hara and J.~E.~Thomas,
  %``Mechanical stability of a strongly interacting Fermi gas of atoms,''
  Phys.\ Rev.\  A {\bf 68}, 011401 (2003).
  %%CITATION = PHRVA,A68,011401;%%
%\cite{Bourdel:2003zz}
\bibitem{Bourdel:2003zz}
  T.~Bourdel, J.~Cubizolles, L.~Khaykovich, K.~M.~F.~Magalhaes, S.~J.~J.~Kokkelmans, G.~V.~Shlyapnikov and C.~Salomon,
  %``Measurement of the Interaction Energy near a Feshbach Resonance in a Li-6
  %Fermi Gas,''
  Phys.\ Rev.\ Lett.\  {\bf 91}, 020402 (2003).
  %%CITATION = PRLTA,91,020402;%%
\bibitem{Gupta:2003} S.~Gupta {\it et al.}, Science {\bf 300}, 1723
(2003).
\bibitem{Regal:2003} C.~A.~Regal and D.~S.~Jin,  Phys.\ Rev.\ Lett.\  {\bf 90}, 230404 (2003)
[arXiv:cond-mat/0302246].

%--------------------------AdS/CFT------------------------

%\cite{hep-th/0104066}
\bibitem{hep-th/0104066} 
  G.~Policastro, D.~T.~Son and A.~O.~Starinets,
  %``The Shear viscosity of strongly coupled N=4 supersymmetric Yang-Mills plasma,''
  Phys.\ Rev.\ Lett.\ \ {\bf 87}, 081601  (2001)
  [hep-th/0104066].
  %%CITATION = PRLTA,87,081601;%%


%------------- bounds on eta/s----------------------

\bibitem{Schafer12}
C.~Chafin, T.~Schafer, [arXiv:1209.1006,[cond-mat.quant-gas]]

\bibitem{Romat12}
P.~Romatschke, R.~E.~Young,  [arXiv:1209.1604,[cond-mat.quant-gas]]

\bibitem{Ku}
M. J. H.~Ku, A. T.~Sommer, L. W.~Cheuk and M. W.~Zwierlein, 
Science {\bf 335}, 563 (2012).


%\cite{Bulgac:2008zz}
\bibitem{Bulgac:2008zz} 
  A.~Bulgac, J.~E.~Drut and P.~Magierski,
  %``Quantum Monte Carlo simulations of the BCS-BEC crossover at finite temperature,''
  Phys.\ Rev.\ A {\bf 78}, 023625 (2008)
  [arXiv:0803.3238 [cond-mat.stat-mech]].
  %%CITATION = ARXIV:0803.3238;%%


%---------------------------Books------------------------

%\bibitem{landaufluids} L. Landau and Lifschitz,
%``Fluid Mechanics'' vol. 6, Prentince Hall, New Jersey.

\bibitem{IntroSupe}
I.~M.~Khalatnikov, ``Introduction to the Theory of Superfluidity",
Benjamin, New York, 1965.




\bibitem{Kosevich}
Yu.A.~Kosevich, Fiz. Nizk. Temp. 9, 479 (1983) [Low Temp. Phys. 9,
242 (1983)].

\bibitem{Maris}
H.~J.~Maris, Phys.\ Rev.\  A {\bf 8}, 1980 (1973).

\bibitem{Benin}
D.~Benin,  Phys.\ Rev.\  B {\bf 11}, 145 (1975).

%---------------------------Effective Lagrangian------------------------
\bibitem{Son:2005rv}
  D.~T.~Son and M.~Wingate,
  %``General coordinate invariance and conformal invariance in nonrelativistic
  %physics: Unitary Fermi gas,''
  Annals Phys.\  {\bf 321}, 197 (2006)
  [arXiv:cond-mat/0509786].
  %%CITATION = APNYA,321,197;%%

\bibitem{valle}
  J.~L.~Ma\~nes and M.~A.~Valle,
Annals of Physics {\bf 324}, 1136 (2009) [arXiv:0810.3797 [cond-mat.other]].

%-------------------our letter ----------------------------
%\cite{Mannarelli:2012su}
\bibitem{Mannarelli:2012su} 
  M.~Mannarelli, C.~Manuel and L.~Tolos,
  %``Shear viscosity in a superfluid cold Fermi gas at unitarity,''
  [arXiv:1201.4006 [cond-mat.quant-gas]].
  %%CITATION = ARXIV:1201.4006;%%
 
%---------------------------c_1 and c_2------------------------
%\cite{Rupak:2008xq}
\bibitem{Rupak:2008xq}
  G.~Rupak and T.~Schafer,
  %``Density Functional Theory for non-relativistic Fermions in the Unitarity Limit,''
  Nucl.\ Phys.\ A\ {\bf 816}, 52  (2009)
  [arXiv:0804.2678 [nucl-th]].
  %%CITATION = ARXIV:0804.2678;%%

\bibitem{Salasnich}
  L.~Salasnich and F.~Toigo,
  %``Viscosity-entropy ratio of the unitary Fermi gas from zero-temperature elementary excitations,''
  J.\ Low.\ Temp.\ Phys.\  {\bf 165}, 239 (2011)
  [arXiv:1107.4552 [cond-mat.quant-gas]].
  %%CITATION = ARXIV:1107.4552;%%


%%---------------------------experiments Duke------------------------
\bibitem{kinast1}  %paper with the data used by us
J.~Kinast, A.~Turlapov and J. E.~Thomas, Phys.\ Rev.\ Lett.\  {\bf 94}, 170404 (2005) [arXiv:cond-mat/0502507].
\bibitem{kinast3}
C.~Cao, E.~Elliott, H.~Wu and  J.E.~Thomas, NJP {\bf 13}, 075007 (2011).

%----------------------- books rarefied gases -----------------------------------
\bibitem{Cercignani}
C.~Cercignani, {\it Rarefied gas dynamics}, Cambridge University Press, Cambridge, 2000
\bibitem{Sone}
Y.~Sone, {\it Kinetic theory and fluid dynamics}, Birkhauser, Boston, 2002
\bibitem{Struchtrup} H.~Struchtrup, 
 {\it Macroscopic transport equations for rarefied gas flows}, Springer, New York, 2005



%\cite{Braby:2010ec}
\bibitem{Braby:2010ec} 
 M.~Braby, J.~Chao and T.~Schafer,
 ``Thermal Conductivity and Sound Attenuation in Dilute Atomic Fermi Gases,''
  Phys.\ Rev.\ A {\bf 82}, 033619 (2010)
  [arXiv:1003.2601 [cond-mat.quant-gas]].
  %%CITATION = ARXIV:1003.2601;%%
  

%\cite{Escobedo:2009bh}
\bibitem{Escobedo:2009bh}
 M.~A.~Escobedo, M.~Mannarelli and C.~Manuel,
 ``Bulk viscosities for cold Fermi superfluids close to the unitary limit,''
  Phys.\ Rev.\  A {\bf 79}, 063623 (2009)
  [arXiv:0904.3023 [cond-mat.quant-gas]].
  %%CITATION = PHRVA,A79,063623;%%


%\cite{Engelbrecht:1997zz}
\bibitem{Engelbrecht:1997zz} 
  J.~R.~Engelbrecht, M.~Randeria and C.~A.~R.~Sa de Melo,
  %``BCS to Bose crossover: Broken-symmetry state,''
  Phys.\ Rev.\ B {\bf 55}, 15153 (1997).
  %%CITATION = PHRVA,B55,15153;%%


  
  
  %\cite{Gubankova:2008ya}
\bibitem{Gubankova:2008ya} 
  E.~Gubankova, M.~Mannarelli and R.~Sharma,
  %``Collective modes in asymmetric ultracold Fermi systems,''
  Annals Phys.\  {\bf 325}, 1987 (2010)
  [arXiv:0804.0782 [cond-mat.supr-con]];
  %%CITATION = ARXIV:0804.0782;%%
  %\cite{Anglani:2011cw}
%\bibitem{Anglani:2011cw} 
  R.~Anglani, M.~Mannarelli and M.~Ruggieri,
  %``Collective modes in the color flavor locked phase,''
  New J.\ Phys.\  {\bf 13}, 055002 (2011)
  [arXiv:1101.4277 [hep-ph]].
  %%CITATION = ARXIV:1101.4277;%%
  
  \bibitem{Diener:2008}
  R. B.~Diener, R.~Sensarma and M.~ Randeria,
  Phys.\ Rev.\ A {\bf 77}, 023626 (2008) .
  
  
  \bibitem{Salasnich-2}
  L.~Salasnich and F.~Toigo, Phys.\ Rev.\ A {\bf 78}, 053626 (2008); Phys. Rev. A 82, 059902(E) (2010

%\cite{Combescot:2006zz}
\bibitem{Combescot:2006zz} 
  R.~Combescot, M.~Y.~.Kagan and S.~Stringari,
  %``Collective mode of homogeneous superfluid Fermi gases in the BEC-BCS crossover,''
  Phys.\ Rev.\ A {\bf 74}, 042717 (2006)
  [cond-mat/0607493 [cond-mat.other]].
  %%CITATION = COND-MAT/0607493;%%



%\cite{Haussmann:2007zz}
\bibitem{Haussmann:2007zz} 
  R.~Haussmann, W.~Rantner, S.~Cerrito and W.~Zwerger,
  %``Thermodynamics of the BCS-BEC crossover,''
  Phys.\ Rev.\ A {\bf 75}, 023610 (2007)
  [cond-mat/0608282 [cond-mat.stat-mech]].
  %%CITATION = COND-MAT/0608282;%%

\bibitem{Haussmann:2008} 
R.~Haussmann and W.~Zwerger,
Phys.\ Rev.\ A {\bf 78}, 063602 (2008)
[arXiv:0805.3226v4 [cond-mat.stat-mech]]. 

\bibitem{Haussmann:2009}
 R.~Haussmann, M.~Punk and W.~Zwerger,
 Phys.\ Rev.\ A {\bf 80}, 063612 (2009) 
[ arXiv:0904.1333v3 [cond-mat.quant-gas]]

%\cite{Arnold:2006fr}
\bibitem{Arnold:2006fr} 
  P.~B.~Arnold, J.~E.~Drut and D.~T.~Son,
  %``Next-to-next-to-leading-order epsilon expansion for a Fermi gas at infinite scattering length,''
  Phys.\ Rev.\ A {\bf 75}, 043605 (2007)
  [cond-mat/0608477 [cond-mat.other]].
  %%CITATION = COND-MAT/0608477;%%
\bibitem{Bertsch}
G.~Bertsch,Proceedings of the10th International
Conference on Recent Progress in Many-Body Theories, R. F.~Bishop et al., Eds. (World Scientific, Singapore, 2000)


%\cite{Escobedo:2010uv}
\bibitem{Escobedo:2010uv}
  M.~A.~Escobedo and C.~Manuel,
  %``Effective field theory and dispersion law of the phonons of a
  %non-relativistic superfluid,''
  Phys.\ Rev.\  A {\bf 82}, 023614 (2010)
  [arXiv:1004.2567 [cond-mat.quant-gas]].
  %%CITATION = PHRVA,A82,023614;%%

%\cite{Manuel:2011ed}
\bibitem{Manuel:2011ed}
  C.~Manuel and L.~Tolos,
  %``Shear viscosity due to phonons in superfluid neutron stars,''
 Phys.\ Rev.\ D {\bf 84},
123007 (2011)  [arXiv:1110.0669 [astro-ph.SR]].
  %%CITATION = ARXIV:1110.0669;%%

\bibitem{Zadorozhko} A.~A.~Zadorozhko,  
{\'E}.~Y.~Rudavski{\u i},  V.~K.~Chagovets,  G.~A.~Sheshin,  
and  Y.~A.~Kitsenko, Low Temperature Physics, {\bf 35},  100 (2009).  
  
  %---------------------------Calculation of eta ------------------------
  \bibitem{Rupak:2007vp}
  G.~Rupak and T.~Schafer,
  %``Shear viscosity of a superfluid Fermi gas in the unitarity limit,''
  Phys.\ Rev.\  A {\bf 76},  053607 (2007)
  [arXiv:0707.1520 [cond-mat.other]].
 %%CITATION = PHRVA,A76,053607;%%
  
   
  
%-----------------------bounds on the transport coefficients---------------------
\bibitem{Hojgaard2}
J.~Hojgaard et al, Phys.\ Rev.\ {\bf 185}, 323 (1969).


%%---------------------------measure of Feshbach position------------------------


%\bibitem{Luo} %paper with the data used by schafer
%L.~Luo and J. E.~Thomas, J. Low Temperature Physics, {\bf 154}, 1  (2009), [arXiv:0811.1159[cond-mat.other]].

\bibitem{Bartenstein}
M.~Bartenstein, A.~Altmeyer, S.~Riedl, R.~Geursen, S.~Jochim, C.~Chin, J. ~Hecker Denschlag, R.`Grimm, A.~Simoni, E.~Tiesinga, C. J.~Williams and P. S.~ Julienne,   Phys.\ Rev.\ Lett.\  {\bf 94}, 103201 (2005).
  
   %--------------experimental measure of eta/s for the cold Fermi gas---------
  %\cite{Cao:2010wa}
\bibitem{Cao:2010wa} 
  C.~Cao, E.~Elliott, J.~Joseph, H.~Wu, J.~Petricka, T.~Schafer and J.~E.~Thomas,
  %``Universal Quantum Viscosity in a Unitary Fermi Gas,''
  Science {\bf 331}, 58 (2011)
  [arXiv:1007.2625 [cond-mat.quant-gas]].
  %%CITATION = ARXIV:1007.2625;%%
  

\bibitem{aboveTc}
G.~M.~Bruun and H.~Smith, 
 Phys.\ Rev.\  A {\bf 75}, 043612 (2007).

\bibitem{Drut}
G.~Wlazlowski, P.~Magierski and J.~E.~Drut, 
 Phys.\ Rev.\ Lett.\  {\bf 109}, 020406 (2012)

\bibitem{dusling}
K.~Dusling and T.~Schaefer, [arXiv:1207.5068[cond-mat.quant-gas]].

\bibitem{Pethick:1975}
C. J.~Pethick, H.~Smith and P.~Bhattacharyya,
Phys.\ Rev.\ Lett.\  {\bf 34}, 643 (1975).
 
 
  %\cite{Guo:2010dc}
\bibitem{Guo:2010dc} 
  H.~Guo, D.~Wulin, C.~-C.~Chien and K.~Levin,
  %``Microscopic Approach to Shear Viscosities in Superfluid Gases: From BCS to BEC,''
  Phys.\ Rev.\ Lett.\  {\bf 107}, 020403 (2011)
  [arXiv:1008.0423 [cond-mat.quant-gas]]; ibid,  New. J. Phys. 13, 075011 (2011))
   [arXiv:1009.4678 [cond-mat.supr-con]
      %%CITATION = ARXIV:1008.0423;%%
  
  %---------------------------Ballistic (Knudsen) regime------------------------
\bibitem{Hojgaard}
J.~Hojgaard et al, J. Low Temperature Physics, {\bf 41},  473 (1980).

\bibitem{Eselson} B.~N.~Esel'Son, 
O.~S.~Nosovitskaya, L.~A.~Pogorelov, 
and  V.~I.~Sobolev,  ZhETF Pis ma Redaktsiiu, {\bf 31}, 34 (1980). 

  
  \bibitem{Niemetz} M.~Niemetz and W.~Schoepe, J. Low Temperature Physics, {\bf 135},  447 (2004).
  
 
\bibitem{Ashcroft-Mermin}
N.W.~Ashcroft, D.~Mermin, {\it Solid State Physics}, Harcourt College Publisher, USA, 1976. 
  





%-----------------------------------------------------------


  
  

 
 



%%---------------------------experiments Duke------------------------

%\bibitem{kinast2}
%J.~Kinast, A.~Turlapov, J.E.~Thomas, Q.~Chen, J.~Stajic, 
%and K.~Levin, Science {\bf 307}, 1296 (2005).


%\bibitem{antiphonons}
%H.~Guo, D.~Wulin, C.-C.~Chien, and K.~Levin,    Phys.\ Rev.\ Lett.\  {\bf 107}, 020403 (2011).  

%
%%-----------------------------dots------------
%\bibitem{schafer-dots}
%T.~Schafer,
% Phys.\ Rev.\ A {\bf 76}, 063618 (2007);   T.~Schafer,
%  %``Elliptic flow and nearly perfect fluidity in dilute Fermi gases,''
%  AIP Conf.\ Proc.\ \ {\bf 1343}, 105  (2011)
%  [arXiv:1012.3718 [nucl-th]].




%\bibitem{SlipReview}
%D.~Einzel and J.P. Parpia, 
 %J.\ Low.\ Temp.\ Phys.\  {\bf 109}, 1 (2007)






\end{thebibliography}
\end{document}